# An iterative data-driven turbulence modeling framework based on Reynolds stress representation

Yuhui Yin[a,b], Yufei Zhang[a, *], Haixin Chen[a], Song Fu[a]

a. School of Aerospace Engineering, Tsinghua University, Beijing 100084, China
b. China Academy of Launch Vehicle Technology, Beijing 100076, China

*\***Corresponding author:** zhangyufei@tsinghua.edu.cn (Yufei Zhang)

[Abstract]
Data-driven turbulence modeling studies have reached such a stage that the basic framework is settled, but several essential issues remain that strongly affect the performance. Two problems are studied in the current research: (1) the processing of the Reynolds stress tensor and (2) the coupling method between the machine learning model and flow solver. For the Reynolds stress processing issue, we perform the theoretical derivation to extend the relevant tensor arguments of Reynolds stress. Then, the tensor representation theorem is employed to give the complete irreducible invariants and integrity basis. An adaptive regularization term is employed to enhance the representation performance. For the coupling issue, an iterative coupling framework with consistent convergence is proposed and then applied to a canonical separated flow. The results have high consistency with the direct numerical simulation true values, which proves the validity of the current approach.

[Key words]
Turbulence modeling, Reynolds-averaged Navier-Stokes equations, Reynolds stress representation, Machine learning

[Highlights]
- Extended the relevant tensor arguments of Reynolds stress by performing the theoretical derivation.
- Proposed an adaptive regularization term to enhance the Reynolds stress representation performance.
- Constructed an iterative coupling framework of the ML model and CFD solver with consistent convergence.

## I. Introduction

Modern engineering design requires high accuracy of flow separation prediction for the computational fluid dynamics (CFD) computation. For the complex turbulent flow separation problem, the traditional turbulence simulation methods either produce unsatisfactory flow prediction or require large computational costs, which cannot meet the requirements of accuracy and efficiency. With the rapid development of data science and machine learning (ML) techniques, the influences of flow structure and physical features that are ignored or difficult to consider in traditional turbulence modeling can be extracted and mapped to the turbulence quantities. The obtained augmented turbulence is referred to as data-driven turbulence modeling. Duraisamy et al.[1] demonstrated the data-driven turbulence modeling as a mathematical expression. First, the baseline turbulence model can be expressed as $\mathcal{M}$:

$$\mathcal{M}\big(\mathbf{w}; \mathcal{P}(\mathbf{w}); \mathbf{c}\big) \tag{1}$$

There exist three elements in the expression. (1) $\mathbf{w}$ represents a set of independent variables selected from mean flow quantities. (2) $\mathcal{P}(\cdot)$ represents the algebraic or differential equations that $\mathbf{w}$ follows. (3) $c$ represents a set of parameters that are generally calibrated with canonical flows.

The high-fidelity data, denoted by $\boldsymbol{\theta}$, can take effect in several aspects to augment the baseline turbulence model, leading to the data-driven turbulence model, $\tilde{\mathcal{M}}$:

$$\tilde{\mathcal{M}}\big(\mathbf{w}(\boldsymbol{\theta}); \mathcal{P}(\mathbf{w},\boldsymbol{\theta}); \mathbf{c}(\boldsymbol{\theta}); \delta(\boldsymbol{\theta},\boldsymbol{\eta})\big) \tag{2}$$

The expression shows four roles that the data can play. (1) Extend the set of independent variables $\mathbf{w}$[2,3]. (2) Modify certain terms in the governing equations $\mathcal{P}(\cdot)$[4–8]. (3) Recalibrate model parameters $\mathbf{c}$[9–12]. (4) Directly model the discrepancy $\delta$ between the model and true values[13–25]. Sometimes the baseline model prediction is neglected, and the discrepancy changes to the entire true value. This situation is also included in the fourth direction.

Different choices of correction terms reflect the researchers' view of where the main discrepancy is located and correspond to different upper limits of augmentation. Regardless of the direction, the final obtained model can be regarded as a new constitutive relation that can predict Reynolds stress and mean flow quantities closer to the true values.

### A. Classification of predicting targets

The complexity of the Reynolds stress leads to different strategies for selecting the predicting targets, each having its advantages and limitations. Four commonly employed choices are separately reviewed as follows from the aspects of spatial invariance, physical interpretability, and smoothness.

(1) The first choice is the multiplying factor of the turbulence model equation term, marked as $\beta(\mathbf{x})$[4–8,26–30]. As a scalar field, predicting $\beta$ can ensure spatial invariance. By multiplying a spatially non-uniform factor field $\beta(\mathbf{x})$, the turbulent quantities in certain areas are altered and the overall performance of the turbulence model can be enhanced. The statistical inference method is generally employed to solve the inverse problem. After acquiring the $\beta$ field, a machine learning model is trained to predict the $\beta$ field using mean flow features $\boldsymbol{\eta}$. The entire process is named field inversion machine learning (FIML)[6–8]. However, the inferred $\beta$ field can yield a quite close result, but the Reynolds stress might be far from the truth because there might exist many different turbulence fields all producing the same mean flow field. The inference cannot guarantee the correct turbulence field.

(2) The second choice is the eddy viscosity $v_t$. Also, as a scalar, predicting $v_t$ satisfies the spatial invariance. A similar problem as $\beta$ is how to acquire the "correct" $v_t$. There are two methods. The first is employing statistical inference. By changing the inferred variable from the turbulence model term to the turbulence model result, the optimal $v_t$ field can be directly acquired[31]. However, the inferred $v_t$ faces the same problem as FIML. The second is to compute the optimal viscosity using the pointwise least-square approximation[3,21]. However, the result might lose clear physical implications and smoothness in the complex flow region where the Reynolds stress anisotropy is remarkable, which deteriorates the model performance.

(3) The third choice involves the Reynolds stress eigenvalues ($k$, $\lambda_1$, $\lambda_2$) and eigenvectors ($\mathbf{v}_1$, $\mathbf{v}_2$, $\mathbf{v}_3$)[2,3,32]. Selecting such targets means discarding all the assumptions and modeling the entire stress as a second-order symmetric tensor. These features can be computed by eigendecomposition of the true value from high fidelity databases, or by inference from observed mean flow quantities[33].

The eigendecomposition method needs to deal with the spatial invariance problem. The invariance of three scalars ($k$, $\lambda_1$, $\lambda_2$) can be guaranteed but the vectors ($\mathbf{v}_1$, $\mathbf{v}_2$, $\mathbf{v}_3$) are naturally spatially variant. One solution is to introduce the baseline eigenvectors and change the targets to the discrepancy between two sets of vectors[3,32]. Such a treatment introduces spatial rotation invariance, but reflection invariance is still missing. In addition, the rotation angle faces discontinuity because of the switching of the eigenvalue ranking and needs further numerical treatment[32].

(4) The fourth choice is the Reynolds stress representation based on the tensor function representation theorem. This method comes from the nonlinear eddy viscosity model (marked as NEVM below) in traditional turbulence modeling. Researchers supposed the Reynolds stress as a tensor function of strain rate $\mathbf{S}$ and rotation rate $\boldsymbol{\Omega}$, which is:

$$\boldsymbol{\tau} = \mathbf{f}(\mathbf{S}, \boldsymbol{\Omega}) \tag{3}$$

In addition, prior physics of turbulence modeling require the tensor function to be isotropic under the extended Galilean transformation[34], which means that the symmetric transformation group of the Reynolds stress function is the entire full orthogonal group (rotation and reflection), which is:

$$\mathbf{Q} \cdot \boldsymbol{\tau}(\mathbf{S}, \boldsymbol{\Omega}) \cdot \mathbf{Q}^{\mathrm{T}} = \boldsymbol{\tau}(\mathbf{Q} \cdot \mathbf{S} \cdot \mathbf{Q}^{\mathrm{T}}, \mathbf{Q} \cdot \boldsymbol{\Omega} \cdot \mathbf{Q}^{\mathrm{T}}) \tag{4}$$

where $\mathbf{Q}$ is a temporal-constant orthogonal matrix.

Pope[35] deduced 10 tensor bases (referred to as the integrity basis) and 5 invariants formed by $\mathbf{S}$ and $\boldsymbol{\Omega}$ using the Cayley-Hamilton theorem. Any symmetric isotropic tensor can be obtained by the linear combination of the 10 tensor bases and the coefficients are functions of the 5 invariants.

One clear advantage of the Reynolds stress representation is a combination of the accuracy and realizability. Taking multiple tensor bases evades the poor performance of $v_t^m$. Meanwhile, the coefficients are all scalars, naturally guaranteeing spatial invariance. Therefore, current research follows this direction.

### B. The coupling method: frozen and iterative

After the predicting targets are selected, we need to select a coupling method between the ML model and the CFD solver, which can be divided into two categories: frozen substitution and iterative substitution.

In frozen substitution, the ML model establishes the mapping from the mean flow features computed by the baseline model to the Reynolds stress true value. Therefore, when used for prediction, the baseline Reynolds-averaged Navier-Stokes (RANS) simulation is performed to acquire the baseline mean features. The model is executed once, and the predicted value corresponds to the true stress. The stress is then substituted into the RANS equations and frozen until convergence. The flow chart is shown in **Fig. 1** (a).

In the iterative substitution, the model establishes the mapping from the true mean flow features obtained from the high-fidelity database, e.g. the direct numerical simulation (DNS) results, to the true Reynolds stress. When used for prediction, using the baseline input features leads to incorrect stress at the beginning. Therefore, iteration is needed, and the model is executed in each CFD iteration step. After the iteration converges, the mean flow and the Reynolds stress both converge to correct results. The flow chart is shown in **Fig. 1** (b).

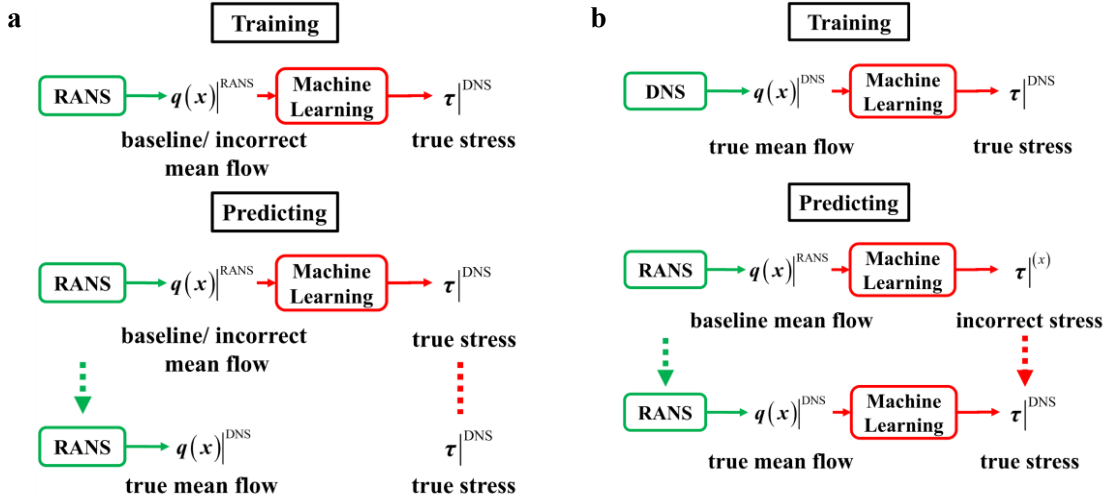

**Fig. 1** Two coupling methods of the ML and CFD solver. **a** The frozen method. **b** The iterative method.

The model training and performance are greatly affected by the coupling method. First, the coupling directly affects the training data preparation. In the frozen framework, the input features are constructed from the baseline model, and the relevant quantities including the primitive variables ($\rho$, **u**, $p$) and turbulence variables ($k$, $\omega$) are easy to acquire. However, in the iterative framework, input features are constructed from the true value, e.g. the DNS result. However, most DNS databases do not provide turbulent dissipation. How to generate a ($\omega$) field compatible with the true result is worth considering. Noted that similar problems are also discussed in several researches by Sandberg et al.[36], Schmelzer et al.[17], Liu et al.[21], and Duraisamy[37]. Given that there is no unified statement of the issue, the problem is referred to as "truth compatibility" in the current research.

Second, the coupling affects convergence and accuracy. Inappropriate Reynolds stress treatments may cause ill-conditioned problems[38], which means that the computed mean flow given the true stress is different from the true mean flow. Relevant studies reached a consensus that the iterative framework has better accuracy and decreases the propagation error. However, the iterative framework may encounter convergence difficulty because the model needs to take the intermediate flow quantities as input and ensure that the iteration ends in the final correct results, which requires the ML model to have strong dynamic robustness.

Third, importantly, the coupling method determines the physical relevance between the input and the stress. In the frozen framework, the mapping from baseline mean flow to true stress lacks sufficient physical implications. The excavated rule by the ML is more like identifying the baseline error region and modifying it, but not a constitutive relation. In contrast, in the iterative framework, the mapping is established from true mean flow to true stress. The physical relevance is more reasonable, and the excavated rule is closer to the nature of turbulence.

In summary, the two coupling methods each have their advantages and limitations. In the current research, since we have selected the Reynolds representation coefficients as targets, the consistency between the mean flow and the stress is more necessary and should be guaranteed first. Therefore, we selected the iterative framework.

### C. Motivation

In the present work, we construct an iterative data-driven turbulence modeling framework based on Reynolds stress representation. Two main processes have been developed. First, we review the selection of tensor arguments that the Reynolds stress depends on and further discuss the physical implications. We then reformulate the tensor invariants and integrity basis using the tensor function representation theorem. The results under two-dimensional flow and three-dimensional flow are given out separately.

Second, a novel iterative framework is proposed. The framework is designed to manifest "consistent convergence" including the truth compatibility and the dynamic robustness mentioned above. In addition, the framework separates the ML process and the Reynolds stress representation, while nearly all earlier studies combined them. The treatment greatly enhances the physical interpretability and smoothness of the coefficients and the final prediction performance.

The rest of this paper is organized as follows. Section II introduces the methodology from three aspects: tensor representation analysis, framework construction, and representation coefficient computation. Section III presents the numerical results of ML model training and prediction of the canonical periodic hill flow. Section IV discusses the effect of the remaining part after the Reynolds stress representation and the ML model calling frequency. Section V summarizes the paper.

## II. Methodology

The three-dimensional compressible RANS equations for a Newtonian fluid without body force, heat transfer, and heat generation are

$$\frac{\partial \rho}{\partial t} + \nabla \cdot (\rho \mathbf{u}) = 0$$

$$\frac{\partial}{\partial t}(\rho \mathbf{u}) + \nabla \cdot (\rho \mathbf{u}\mathbf{u}) = \nabla \cdot \mathbf{T}$$

$$\frac{\partial}{\partial t}(\rho e) + \nabla \cdot (\rho e \mathbf{u}) = \nabla \cdot (\mathbf{T} \cdot \mathbf{u})$$

$$\mathbf{T} = -p\mathbf{I} + \left[\lambda (\nabla \cdot \mathbf{u})\mathbf{I} + \mu (\nabla \mathbf{u} + \mathbf{u}\nabla)\right] + \boldsymbol{\tau}$$

(5)

where $\lambda$ and $\mu$ are the bulk viscosity and molecular viscosity, respectively. $\mathbf{T}$ represents the total stress tensor, including the pressure, the molecular viscous stress, and the Reynolds stress $\boldsymbol{\tau}$ that must be closed.

Regardless of traditional modeling or data-driven modeling, $\boldsymbol{\tau}$ can always be expressed as follows:

$$\boldsymbol{\tau} = \mathbf{f}(\mathbf{S}, \boldsymbol{\Omega}, \cdots; T_{\text{turb}}, L_{\text{turb}})$$

(6)

$(\mathbf{S}, \boldsymbol{\Omega}, \cdots)$ represents tensor arguments of the Reynolds stress function. The argument can be a vector, two-order symmetric tensor, or two-order antisymmetric tensor. $(T_{\text{turb}}, L_{\text{turb}})$ represents the turbulent time scale and length scale, which are generally constructed by ($k$, $\omega$). To build a new constitutive relation, the first step is to select proper arguments. Then the tensor invariants and integrity basis can be acquired, which leads to the first part of the methodology.

### A. Tensor representation analysis

#### 1. Extension of the tensor arguments

Tensor representation analysis has been employed in turbulence modeling for decades. As mentioned above, since Lumley[39] and Pope[35] proposed a complete form of the NEVM, with the development of subsequent studies[40–42], the model using $(\mathbf{S}, \boldsymbol{\Omega})$ as tensor arguments is currently being perfected. In the current research, we start from the basic nonlinear eddy viscosity model and explore other potential tensor arguments apart from $(\mathbf{S}, \boldsymbol{\Omega})$. More specifically, we evaluate the hypotheses of the NEVM and introduce additional quantities when these hypotheses do not hold.

The original form of the NEVM deduced by Lumley[39] is as follows:

$$\boldsymbol{\tau} = q^2 \mathbf{h}\{\mathbf{S}(\boldsymbol{\xi}), \boldsymbol{\Omega}(\boldsymbol{\xi})\}$$

(7)

where $q$ is the turbulent velocity scale, and $\boldsymbol{\xi} = \mathbf{x}/(q^3/\varepsilon)$ is the nondimensional spatial coordinate normalized by $q$ and the dissipation rate $\varepsilon$. Three assumptions are used during the deduction:

(1) incompressible turbulence, (2) weak historical effect.

The first assumption is the incompressible hypothesis. In the incompressible flow, the pressure can be obtained from the mean velocity field through the Poisson equation, and the velocity field contains all the mean field information. However, most flow problems in actual engineering are compressible. The pressure becomes an independent state variable, which should be added to the arguments set.

The second assumption is the weak historical effect and homogeneous hypothesis. The consideration of including the historical effect and inhomogeneous turbulence is also a key direction in traditional modeling research. A commonly used approach is to consider the temporal and spatial historical effects in $\mathbf{S}$ to produce an "effective" strain rate $\tilde{\mathbf{S}}$ [43], which can be expressed as a convolution form:

$$\tilde{\mathbf{S}}(t) = \int_{-\infty}^{t} \mathbf{S} \frac{e^{-(t-\tau)/\Lambda_m(t)}}{\Lambda_m(t)} D\tau \tag{8}$$

where $\Lambda_m$ is the turbulent time scale. We perform series expansion at local ($\mathbf{x}$, $t$) on the equation above:

$$\tilde{\mathbf{S}} = \mathbf{S} + \sum_{n=1}^{\infty} (-\Lambda_m)^n \left[ \frac{D^n \mathbf{S}}{Dt^n} - \frac{\mathbf{I}}{3} \text{tr}\left( \frac{D^n \mathbf{S}}{Dt^n} \right) \right] \tag{9}$$

Taking the first-order approximation, the above expression shows that $\tilde{\mathbf{S}}$ includes the local $\mathbf{S}$ and the total derivative $D\mathbf{S}/Dt$. We further deduce the transport equation of $\mathbf{S}$ by applying the left gradient and right gradient to the mean velocity equation and summing them. The final result is shown below:

$$\frac{D\mathbf{S}}{Dt} = -\left(\mathbf{S}^2 + \mathbf{\Omega}^2\right) + \nu \Delta \mathbf{S} - \nabla \nabla \left( \frac{p}{\rho} \right) - \frac{1}{2} \left( \nabla \nabla \cdot \mathbf{\tau} + \mathbf{\tau} \cdot (\nabla \nabla) \right) \tag{10}$$

We analyze the right-hand side of the equation in sequence. $(\mathbf{S}^2+\mathbf{\Omega}^2)$ can be expressed by the integrity basis of ($\mathbf{S}$, $\mathbf{\Omega}$). $\nu \Delta \mathbf{S}$ represents the viscous diffusion and is not an active source term dominating the historical effect. The remaining two terms are the pressure gradient related term and the Reynolds stress gradient related term. If we want to represent the effect of $\tilde{\mathbf{S}}$, the two gradients should be included.

Based on the analysis above, this paper adds two additional tensor arguments into the original ($\mathbf{S}$, $\mathbf{\Omega}$), which are the pressure gradient vector $\mathbf{v}_p$ and the turbulent kinetic energy

(marked as TKE below) gradient vector $\mathbf{v}_k$, defined as follows:

$$\mathbf{v}_p = \nabla(p/\rho) \quad \mathbf{v}_k = \nabla k \tag{11}$$

The final Reynolds stress isotropic tensor function is

$$\boldsymbol{\tau} = \mathbf{f}\left(\mathbf{S},\boldsymbol{\Omega},\mathbf{v}_p,\mathbf{v}_k;T_{\text{turb}},L_{\text{turb}}\right) \tag{12}$$

Generally, the TKE is solved by its own transport equation and the expression above can be normalized to

$$\boldsymbol{\tau} = 2k\left(\frac{\mathbf{I}}{3} + \mathbf{b}\left(\hat{\mathbf{S}},\hat{\boldsymbol{\Omega}},\hat{\mathbf{v}}_p,\hat{\mathbf{v}}_k\right)\right) \tag{13}$$

where $\mathbf{b}$ is the nondimensional Reynolds deviatoric tensor $\mathbf{b} = \boldsymbol{\tau}/2k - \mathbf{I}/3$, and the superscript $(\hat{\cdot})$ means normalization using turbulence scales, as equation (14) shows.

$$\hat{\mathbf{S}} = \frac{\mathbf{S}}{\omega} \quad \hat{\boldsymbol{\Omega}} = \frac{\boldsymbol{\Omega}}{\omega} \quad \hat{\mathbf{v}}_p = \frac{\mathbf{v}_p}{\omega\sqrt{k}} \quad \hat{\mathbf{v}}_k = \frac{\mathbf{v}_k}{\omega\sqrt{k}} \tag{14}$$

It is worth mentioning that the final tensor argument set is basically the same as in earlier studies[3,32], except including the density into the pressure term and the alternative normalization. The main purpose of this part is to systematically deduce the additional tensor arguments rather than determine the arguments ad hoc.

**2. Complete irreducible tensor invariants and integrity basis**

The implication of the tensor function representation is briefly introduced in Section I A. Here, we restate the representation of $\mathbf{b}$ in a mathematical manner. Given a set of tensor arguments, the isotropic tensor function representation theorem indicates that any composed tensor function can be expressed as a linear combination of several tensor bases:

$$\mathbf{b} = \sum_{i=1}^{w} g_i\left(I_1 \sim I_a\right)\mathbf{T}_i \tag{15}$$

$(I_1 \sim I_a)$ are the complete and irreducible tensor invariants formed by the argument set. Completeness means that all the other invariants can be represented by these invariants. Irreducibility means that they are independent of each other. $\mathbf{T}_i$, $i=1\sim w$ are the complete and irreducible tensor bases. $\mathbf{T}_i$ are all second-order symmetric tensors and collectively referred to as the integrity basis. $g_i$ are the representation coefficients corresponding to the tensor bases, which are all functions of the invariants ($I_1 \sim I_a$).

The representation process is to acquire the invariants and integrity basis. Theoretically this set of invariants and tensor bases is applicable to all symmetric tensors. If identifying **b** as the target, a corresponding set of $g_i$ is settled simultaneously.

There are two methods to compute the invariants and integrity basis. The traditional method uses the Cayley-Hamilton theorem. This theorem indicates that a high-degree tensor polynomial can be expressed by low-degree polynomials. In the actual computation process, the general form of the tensor polynomial composed of this set of tensor arguments is given first. Then, the C-H theorem is repeatedly applied to the polynomial to simplify the expression. Finally, a set of low-degree tensor bases and invariants appearing in the reduction process are acquired. The deduction of Pope[35] followed this method. However, there are deficiencies. First, the general form of tensor polynomials is easy to acquire only if the number of arguments is small, for example, only **S** and **Ω**. If the number increases, the general form will be complex and multiple, increasing the difficulty of reduction. Second, it is also difficult to prove whether the final results are irreducible. For example, the 10 tensor bases given by Pope were proven not to be the minimal representation in later research[40].

We employ the other method proposed by Zheng[44] which directly constructs the basis rather than simplifying it from a complex situation. This method can directly and systematically deduce the results for any number of tensor arguments. We suppose a symmetric tensor function **H** is composed of $L$ symmetric tensors $\mathbf{A}_i$, $m$ antisymmetric tensors $\mathbf{W}_p$, and $N$ vectors $\mathbf{v}_m$:

$$\mathbf{H}\left(\mathbf{A}_i, \mathbf{W}_p, \mathbf{v}_m\right) \tag{16}$$

It is difficult to directly find invariants and tensor bases and to verify their completeness and irreducibility. Therefore, introducing intermediate variables to transform the problem is necessary. For the construction of invariants, the intermediate variable is each component of tensor arguments. For the construction of tensor bases, the intermediate variable is the complete orthogonal basis of the symmetric second-order tensor in space.

For invariants, because of the definition of invariance, they can be computed using the components of all arguments no matter in which coordinate system and remain unchanged. Therefore, in turn, if we construct a set of invariants that can represent all the components in a certain coordinate, these invariants are complete and can represent all the scalar-valued functions. The requirements above for the invariants ($I_1 \sim I_a$) can be described mathematically as:

$$\chi_t = f_t(I_1, \cdots, I_a), \quad t = 1, \cdots, 6L + 3M + 3N \tag{17}$$

where $\chi_t$ represents each component of tensor arguments. Because a symmetric tensor contains 6 independent components, an antisymmetric tensor contains 3, and a vector contains 3, there are in total ($6L+3M+3N$) components.

For tensor bases, after acquiring the invariants above, to further represent a symmetric tensor-valued function, we should select a set of tensor bases that can express all 6 complete orthogonal bases expanding the entire symmetric tensor space, which is described as:

$$\begin{aligned} \mathbf{E}_1 &= \sum_{w=1}^{c} \eta_w^{(1)} \mathbf{T}_w \\ &\cdots \\ \mathbf{E}_6 &= \sum_{w=1}^{c} \eta_w^{(6)} \mathbf{T}_w \end{aligned} \tag{18}$$

where the coefficients $\eta_w$ are isotropic functions of ($I_1 \sim I_a$) and $\mathbf{E}_1$ - $\mathbf{E}_6$ are:

$$\begin{aligned} &\mathbf{E}_1 = \mathbf{e}_1 \otimes \mathbf{e}_1 && \mathbf{E}_2 = \mathbf{e}_2 \otimes \mathbf{e}_2 && \mathbf{E}_3 = \mathbf{e}_3 \otimes \mathbf{e}_3 \\ &\mathbf{E}_4 = \mathbf{e}_2 \otimes \mathbf{e}_3 + \mathbf{e}_3 \otimes \mathbf{e}_2 && \mathbf{E}_5 = \mathbf{e}_3 \otimes \mathbf{e}_1 + \mathbf{e}_1 \otimes \mathbf{e}_3 && \mathbf{E}_6 = \mathbf{e}_1 \otimes \mathbf{e}_2 + \mathbf{e}_2 \otimes \mathbf{e}_1 \end{aligned} \tag{19}$$

$\mathbf{e}_1$ - $\mathbf{e}_3$ are three independent orthogonal basis vectors constructing the entire space.

One advantage of the method is that the coordinate can be properly selected to minimize the number of components needing representation. For example, when dealing with a symmetric tensor, the coordinate can be the same as its principal axes, and there only exist 3 independent components. When multiple tensor arguments exist, all the possibilities between the principal axes need to be considered and the final complete form is acquired.

In the current research, we employ the method above and further develop the conclusion. The original work only gave the general expression in three-dimensional space of the situation in which the numbers of $\mathbf{A}_i$, $\mathbf{W}_p$ are arbitrary but all the $\mathbf{v}_m$ are collinear. We deduce the situation with arbitrary numbers of $\mathbf{A}_i$, $\mathbf{W}_p$ and $\mathbf{v}_m$. In addition, the expression in two-dimensional space is also acquired. The two expressions are listed in the Appendix. Here we only give the two-dimensional situation of the argument set $(\hat{\mathbf{S}}, \hat{\mathbf{\Omega}}, \hat{\mathbf{v}}_p, \hat{\mathbf{v}}_k)$. There are a total of 11 invariants and 7 tensor bases, as follows:

$$\text{invariants} \quad \begin{array}{llll} I_1 = \hat{\mathbf{v}}_p \cdot \hat{\mathbf{v}}_p & I_2 = \hat{\mathbf{v}}_k \cdot \hat{\mathbf{v}}_k & I_3 = \hat{\mathbf{v}}_p \cdot \hat{\mathbf{v}}_k & I_4 = \mathrm{tr}\hat{\mathbf{S}}^2 \\ I_5 = \mathrm{tr}\hat{\boldsymbol{\Omega}}^2 & I_6 = \hat{\mathbf{v}}_p \cdot \hat{\mathbf{S}}\hat{\mathbf{v}}_p & I_7 = \hat{\mathbf{v}}_k \cdot \hat{\mathbf{S}}\hat{\mathbf{v}}_k & I_8 = \hat{\mathbf{v}}_p \cdot \hat{\mathbf{S}}\hat{\mathbf{v}}_k \\ I_9 = \hat{\mathbf{v}}_p \cdot \hat{\boldsymbol{\Omega}}\hat{\mathbf{v}}_k & I_{10} = \hat{\mathbf{v}}_p \cdot \hat{\mathbf{S}}\hat{\boldsymbol{\Omega}}\hat{\mathbf{v}}_p & I_{11} = \hat{\mathbf{v}}_k \cdot \hat{\mathbf{S}}\hat{\boldsymbol{\Omega}}\hat{\mathbf{v}}_k & \end{array} \quad (20)$$

$$\begin{array}{l} \text{integrity basis} \\ \text{of } \mathbf{b} \end{array} \quad \begin{array}{llll} \mathbf{T}_1 = \hat{\mathbf{S}} & \mathbf{T}_2 = \hat{\mathbf{S}}\hat{\boldsymbol{\Omega}} - \hat{\boldsymbol{\Omega}}\hat{\mathbf{S}} & \mathbf{T}_3 = \hat{\mathbf{v}}_p \otimes \hat{\mathbf{v}}_p & \mathbf{T}_4 = \hat{\mathbf{v}}_k \otimes \hat{\mathbf{v}}_k \\ \mathbf{T}_5 = \hat{\mathbf{v}}_p \otimes \hat{\mathbf{v}}_k + \hat{\mathbf{v}}_k \otimes \hat{\mathbf{v}}_p & \mathbf{T}_6 = \hat{\mathbf{v}}_p \otimes \hat{\boldsymbol{\Omega}}\hat{\mathbf{v}}_p + \hat{\boldsymbol{\Omega}}\hat{\mathbf{v}}_p \otimes \hat{\mathbf{v}}_p & & \\ \mathbf{T}_7 = \hat{\mathbf{v}}_k \otimes \hat{\boldsymbol{\Omega}}\hat{\mathbf{v}}_k + \hat{\boldsymbol{\Omega}}\hat{\mathbf{v}}_k \otimes \hat{\mathbf{v}}_k & & & \end{array} \quad (21)$$

The form of vectors $v_p$ / $v_k$ is worthy of mention. In earlier studies, the vectors are transformed to corresponding antisymmetric tensors using $\mathbf{A} = -\boldsymbol{\varepsilon} \cdot \mathbf{v}$, where $\boldsymbol{\varepsilon}$ is the permutation tensor. In the current research, the vectors are directly introduced. To verify which treatment is rational, we take $\mathbf{A}_k = -\boldsymbol{\varepsilon} \cdot \mathbf{v}_k$ in a two-dimensional situation as an example. The components of **b**, **S**, and $\mathbf{A}_k$ are as follows:

$$\mathbf{b} = \begin{bmatrix} b_{11} & b_{12} & 0 \\ b_{21} & b_{22} & 0 \\ 0 & 0 & 0 \end{bmatrix}, \mathbf{S} = \begin{bmatrix} S_{11} & S_{12} & 0 \\ S_{21} & S_{22} & 0 \\ 0 & 0 & 0 \end{bmatrix}, \mathbf{A}_k = \begin{bmatrix} 0 & 0 & A_{13} \\ 0 & 0 & A_{23} \\ A_{31} & A_{32} & 0 \end{bmatrix} \quad (22)$$

where $A_{13} = -A_{31} = \partial k / \partial y$, $A_{23} = -A_{32} = -\partial k / \partial x$. According to the representation theorem, if $\mathbf{A}_k$ is used, there should be $(\mathbf{S}\mathbf{A}_k - \mathbf{A}_k\mathbf{S})$ in the integrity basis, the components of $(\mathbf{S}\mathbf{A}_k - \mathbf{A}_k\mathbf{S})$ can be computed as:

$$\mathbf{S}\mathbf{A}_k - \mathbf{A}_k\mathbf{S} = \begin{bmatrix} 0 & 0 & A_{13}S_{11} + A_{23}S_{12} \\ 0 & 0 & A_{13}S_{12} + A_{23}S_{22} \\ A_{13}S_{11} + A_{23}S_{12} & A_{13}S_{12} + A_{23}S_{22} & 0 \end{bmatrix} \quad (23)$$

It can be found that this term makes no contribution to the component to the deviatoric stress **b**. Therefore, it is physically unreasonable. As a comparison, if $\mathbf{v}_k$ is used, the term $(\mathbf{v}_k \otimes \mathbf{v}_k)$ is:

$$\mathbf{v}_k \otimes \mathbf{v}_k = \begin{bmatrix} v_1 v_1 & v_1 v_2 & 0 \\ v_1 v_2 & v_2 v_2 & 0 \\ 0 & 0 & 0 \end{bmatrix} \quad (24)$$

where $v_1 = \partial k / \partial x, v_2 = \partial k / \partial y$. This term makes an effective contribution to **b**.

In summary, in this part, we extend the tensor argument set from (**S**, **Ω**) to (**S**, **Ω**, **v**$_p$, **v**$_k$) and deduce the corresponding invariants and integrity basis in three-dimensional and two-dimensional situations. The difference between the current treatment and earlier studies is analyzed, proving the validity.

<p align="center">**B. Framework construction**</p>

In this part, we illustrate the construction of the data-driven turbulence modeling framework according to the sequence of preprocessing, training, predicting, and solving.

**1. Training data preparation**

The first step of preprocessing is to determine what data should be used for the construction. The flow case used for the training and testing in the current research is the periodic hill. There exists a massive flow separation area near the leeward side of the curved wall, which makes it difficult for the RANS method to accurately simulate the flow. The parameterized periodic hill geometries and corresponding databases provided by Xiao[45] are employed. The factor $α$ is introduced to scale the hill width, as shown in **Fig. 2**. The Reynolds number based on $H$ and the bulk velocity at crest $U_b$ is Re$_b$ = 5,600 for all 5 cases. The DNS simulations are conducted under incompressible conditions. The $α$ = 0.8 and 1.2 are selected as the training set and $α$ = 0.5, 1.0, and 1.5 are selected as the testing set, which is consistent with our previous work[32].

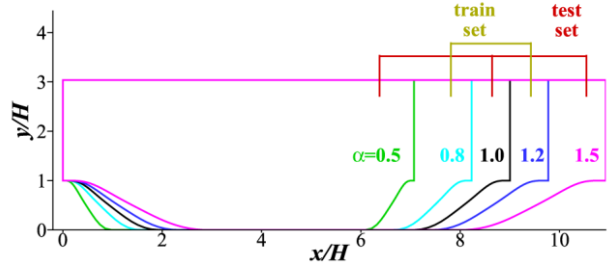

**Fig. 2** Parameterized periodic hill geometries with different $α$[32]

The following issue after the flow case selection is to determine the status in which the training data should be. In Section I B, we propose the truth compatibility requirement, which is simply described as "all the features are constructed from the true value". The subsequent problem is that the true values of some quantities cannot be acquired from the high-fidelity database. To overcome this problem, we extend the implication of truth compatibility by substituting the "true value" with the "end-state value".

The end-state value is defined as the quantity for which the computation converges. To better illustrate the end-state of different variables, the CFD solving process is explained first, as shown in **Fig. 3**. The lateral axis represents the iteration. The left part shows the baseline RANS process. The green and blue arrows represent the iteration of the mean and turbulent flow, respectively. The black arrows represent the data transfer, and the dotted lines represent the converged results. The right part shows the iterative computation. The ML model is introduced and executed in each iteration step (red arrows). The mean flow, ML-predicted Reynolds stress, and turbulent flow each have their own convergence paths and end-states. The mean flow receives the Reynolds stress from the ML model and the end-state value is the DNS true value, which is the ultimate aim of the framework. The ML model computes the Reynolds stress using input features from mean flow and turbulent flow and transfers the stress to the mean flow. The end-state value is also the DNS true value because only the true Reynolds stress corresponds to the true mean flow. However, the turbulent flow follows a different path. During the iteration, $(k, \omega)$ only serves as the reference value to nondimensionalize the input features of the ML model, which means that the turbulence equations only accept mean flow results and no longer transfer back. Therefore, the end-state values of $(k, \omega)$ do not correspond to the true turbulence, but instead the converged results of the turbulence equations given the true mean flow, as shown by the blue dotted line.

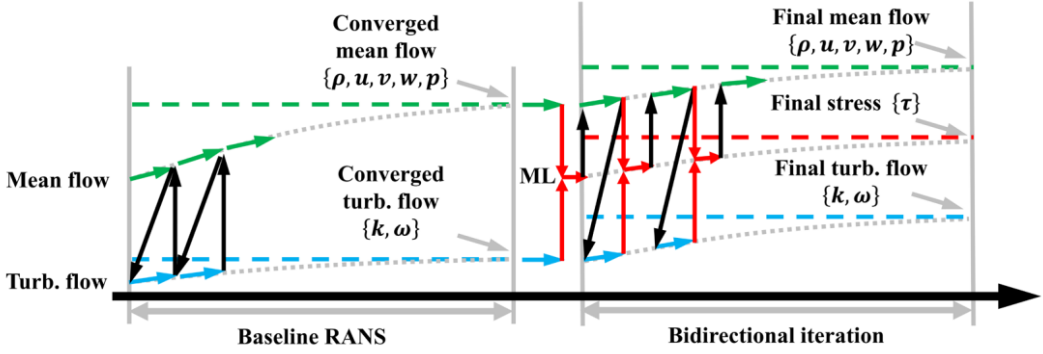

**Fig. 3** CFD solution process of the iterative framework

In summary, the end-state values are divided into two categories. The first type corresponds to the DNS true value, including the mean flow features such as **S**, **Ω**, and $\nabla(p/\rho)$ and the Reynolds stress $\boldsymbol{\tau}$. The training data should be interpolated from the DNS fine grid to the RANS coarse grid and discretized if gradients are needed. The second type corresponds to the variables that are not true but compatible with the true values, including the turbulent flow features given by the baseline turbulence equations $(k, \omega)$. To acquire the results, the turbulence model needs to be solved solely with all the mean flow terms frozen as the DNS true values. Such $(k, \omega)$ are different from both the original baseline RANS and the true DNS and are referred to as pseudo DNS results in this paper, abbreviated as $(k, \omega)^{\text{pDNS}}$. The $v_t$ used in the following features is also computed by $(k, \omega)^{\text{pDNS}}$.

At the end of this section, the truth compatibility problem is further discussed and the differences between the current research and the former researches[17,21,36] are clarified. Computing the frozen turbulence model equations using true mean flow quantities is a commonly employed method in iterative frameworks to acquire the compatible turbulent quantities. The first difference is the treatment of TKE. In the work of Liu et al.[21], the $k^{\text{pDNS}}$ is the only TKE appearing in the framework, which is reasonable because the predicting target in the work is $v_t$ and the true TKE is not necessary. Schmelzer et al.[17] employed a more complicated treatment of TKE by constructing the $P_k$ term with the true Reynolds deviatoric tensor **b** and adding an additional corrective term $R$. The purpose is to directly computing the true TKE with the modified turbulence model equation. The attempt to separating different aspects of correction is fascinating but the left remedy term $R$ still lacks physical interpretability.

The treatment in current research is a trade-off between the two methods above. The original turbulence model equations are not modified but only substituted with true mean flow quantities. The acquired $k^{\text{pDNS}}$ only works as the turbulent scales for nondimensionalization. When predicting the entire Reynolds stress, the TKE is predicted using a sole machine learning model corresponding to the true TKE from DNS results. The machine learning model of TKE will be further illustrated in Section B 3. As a summary, the current method retains the original simple and robust form of the turbulence model equations and predicts the true TKE for the reconstruction of the Reynolds stress simultaneously.

**2. Input feature selection**

Once the training data are acquired, the next step is to construct the input feature set. In this paper we follow the feature selection criteria proposed in our earlier research[32]. The input features are constructed from two perspectives: tensor analysis and flow characteristic.

The tensor analysis perspective is based on the Reynolds representation illustrated above. The acquired 11 complete irreducible tensor invariants of (**S**, **Ω**, $\mathbf{v}_p$, $\mathbf{v}_k$) are shown in equation (20). The number and some expressions of invariants in the current research are different from those in earlier research. This is because a new treatment of vectors is employed, and the new one is more effective as proven in Section II A.

The flow characteristic perspective includes two parts: flow structure identification and turb./mean flow relative strength. The first part extracts the key flow structures relevant to turbulence production or dissipation, such as the free shear layer and the strong adverse pressure gradient flow, by selecting several marker functions. The second part directly shows the intensity distribution of the turbulence, which can help locate the region with high uncertainty and provide the modifying direction.

Considering the smoothness and effectiveness, not all 11 invariants are selected. Those high-degree invariants with small magnitudes and unsmooth distributions are abandoned and 4 invariants $q_1$-$q_4$ remain. The flow characteristic perspective provides 5 more input features $f_1$ - $f_5$. Note that the marker of the shear layer and swirl flow has a different expression compared to the previous work[32] because the formerly used $d^2\Omega/(\nu_t+\nu)$ leads to large values far away from the wall.

Therefore, the final input feature set with 9 features is established and listed in **Table 1**. It is worth mentioning that the invariants employ a different normalization method and are marked with the superscript $\widehat{(\cdot)}$, shown as follows:

$$\widehat{\mathbf{S}} = \frac{\mathbf{S}}{\|\mathbf{S}\|+\omega} \qquad \widehat{\mathbf{\Omega}} = \frac{\mathbf{\Omega}}{\|\mathbf{\Omega}\|+\omega} \qquad \widehat{\mathbf{v}}_p = \frac{\mathbf{v}_p}{\|\mathbf{v}_p\|+\omega\sqrt{k}} \quad \widehat{\mathbf{v}}_k = \frac{\mathbf{v}_k}{\|\mathbf{v}_k\|+\omega\sqrt{k}} \qquad (25)$$

where the symbol "$\|\cdot\|$" represents the tensor norm. Such a "$a/(|a|+b)$" treatment can constrain the value range to (-1, 1) without significantly changing the original distribution. The thought is also employed in some flow characteristic perspective features.

**Table 1:** Input features in the current research

| Feature implication | Expression |
| --- | --- |
| Selected invariants of $(\mathbf{S}, \mathbf{\Omega}, \mathbf{v}_p, \mathbf{v}_k)$ | $q_1 = \mathrm{tr}\widehat{\mathbf{S}}^2$ |
|  | $q_2 = \mathrm{tr}\widehat{\mathbf{\Omega}}^2$ |
|  | $q_3 = \widehat{\mathbf{v}}_p \cdot \widehat{\mathbf{v}}_p$ |
|  | $q_4 = \widehat{\mathbf{v}}_k \cdot \widehat{\mathbf{v}}_k$ |
| Marker of shear layer and swirl flow | $f_1 = \Omega/(\Omega+\omega)$, where $\Omega = \sqrt{2\Omega_{ij}\Omega_{ij}}$ |
| Marker of adverse pressure gradient flow[46] | $f_2 = \dfrac{\widehat{\mathbf{u}}\cdot\nabla\widehat{p}}{1+|\widehat{\mathbf{u}}\cdot\nabla\widehat{p}|}$, where $\widehat{\mathbf{u}} = \mathbf{u}/\sqrt{\mathbf{u}\cdot\mathbf{u}}$ and $\nabla\widehat{p} = \nabla p / \sqrt{(\nabla p)\cdot(\nabla p)}$ |
| Marker of boundary layer[47] | $f_3 = 1 - \tanh\left([8r_d]^3\right)$, where $r_d = (\nu_t+\nu)/\kappa^2 d^2 \sqrt{u_{i,j}u_{i,j}}$ |
| Ratio of turb./mean kinetic energy | $f_4 = 2k/(2k+u_i u_i)$ |
| Ratio of turb./mean viscosity | $f_5 = \nu_t/(\nu_t+\nu)$ |

Four representative features are shown in **Fig. 4**. The white solid lines indicate the flow paths of the DNS results. Different features correspond to different flow structures and regions, such as the boundary layer, shear layer, and adverse pressure gradient flow. A significant advantage is that all features are compatible with the true mean flow rather than the baseline mean flow in the frozen framework, which guarantees a "truth to truth" mapping.

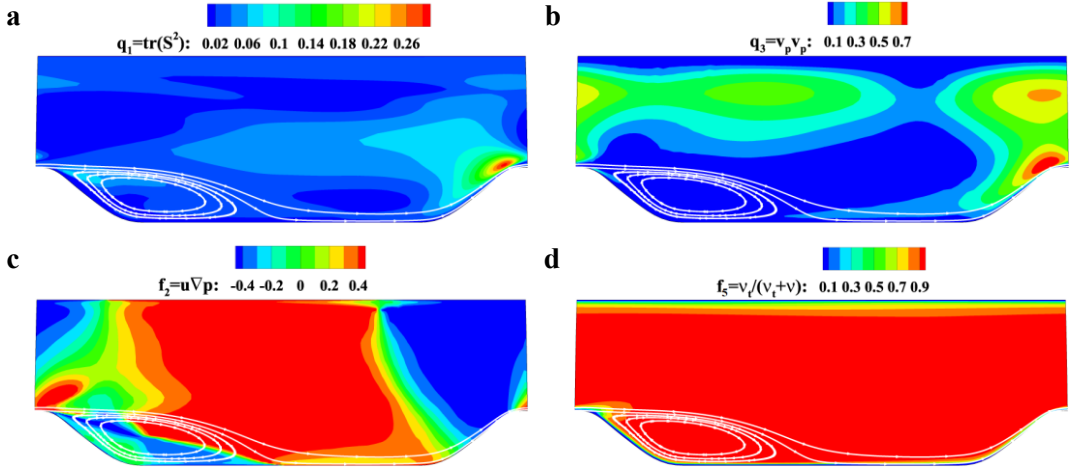

**Fig. 4** Representative input features. **a** $q_1=\mathrm{tr}(\mathbf{S}^2)$. **b** $q_3=\mathbf{v}_p\cdot\mathbf{v}_p$. **c** $f_2=\mathbf{u}\cdot\nabla p$. **d** $f_5=v_t/(v_t+v)$.

### 3. Predicting target selection

We have illustrated the concept of the Reynolds stress representation in Section II A. The representation form is given in equation (15) and the expression of tensor bases is given in equation (21). Therefore, the only unknown variables are the representation coefficients $g_i$. In traditional modeling, the stress expression is substituted into the algebraic stress equation and simplified. The acquired coefficients are complex polynomials of the invariants[40]. In data-driven modeling, with the help of the DNS database, the determination of $g_i$ becomes a numerical optimization problem.

The predicting targets in former ML frameworks based on the Reynolds stress representation generally follow the idea of a tensor basis neural network (marked as TBNN below)[13]. This concept is characterized by embedding the combination of $g_i$ and $\mathbf{T}_i$ into the ML model, as shown in **Fig. 5** (a). The output of the ML model is the directly predicted deviatoric stress $\mathbf{b}^{\mathrm{pred}}$, and the loss function is defined as

$$L = \left\| \mathbf{b}^{\mathrm{true}} - \mathbf{b}^{\mathrm{pred}} \right\| = \left\| \mathbf{b}^{\mathrm{true}} - g_i \mathbf{T}_i \right\| \tag{26}$$

The $g_i$ term is not explicitly shown, but only performs as the latent variables. No more preprocessing is needed besides separating the DNS Reynolds stress $\boldsymbol{\tau}^{\mathrm{true}}$ into the magnitude and $\mathbf{b}^{\mathrm{true}}$.

Several deficiencies exist in the framework above. First, embedding $g_i\mathbf{T}_i$ into the ML model makes $g_i$ inaccessible. The training process only minimizes the discrepancy of the final combination, while the distribution of each coefficient is ignored. This mixes the error of model training and the representation. Second, the estimation of the TKE is generally missing in former studies. The reason for this might be that the TBNN using dimensionless features as input cannot be directly used to map a dimensional quantity.

To overcome these shortcomings, we developed an improved framework and corresponding prediction targets, as shown in **Fig. 5** (b). The framework is separated into two parts, the ML part and the Reynolds stress representation part. The purpose of the representation part is to acquire the true coefficient set $g_i^{\text{true}}$ and the TKE discrepancy $\Delta \ln k$. The $g_i^{\text{true}}$ is determined by the pointwise least-square approximation, as follows:

$$g_i = \arg\min\left(\left\|\mathbf{b} - g_i\mathbf{T}_i\right\|^2\right) \tag{27}$$

The TKE discrepancy is defined as the logarithm of the ratio between the true value and the pDNS value, as follows:

$$\Delta \ln k = \ln\left(k^{\text{true}} / k^{\text{pDNS}}\right) \tag{28}$$

After representation, the ML model is trained to model $g_i^{\text{true}}$ and $\Delta \ln k$.

The current framework can remedy the problems of the former framework in the following aspects. First, the true TKE value is predicted, which further improves the modeling accuracy of the Reynolds stress. The treatment of $\Delta \ln k$ makes the target dimensionless and suitable for the ML model. In addition, by introducing the pDNS value computed by the baseline turbulence model, prior knowledge is included in the model, which can also decrease the modeling difficulty and benefit the convergence. Second, the separation of the ML and the Reynolds stress representation enables us to evaluate the physical interpretation and smoothness of representation coefficients before the training. As will be stated in Section II C, the direct computation of $g_i^{\text{true}}$ using equation (27) yields severely nonphysical and unsmooth distributions. In a separated representation part, the constraints and regularization are easier to impose without affecting the model training.

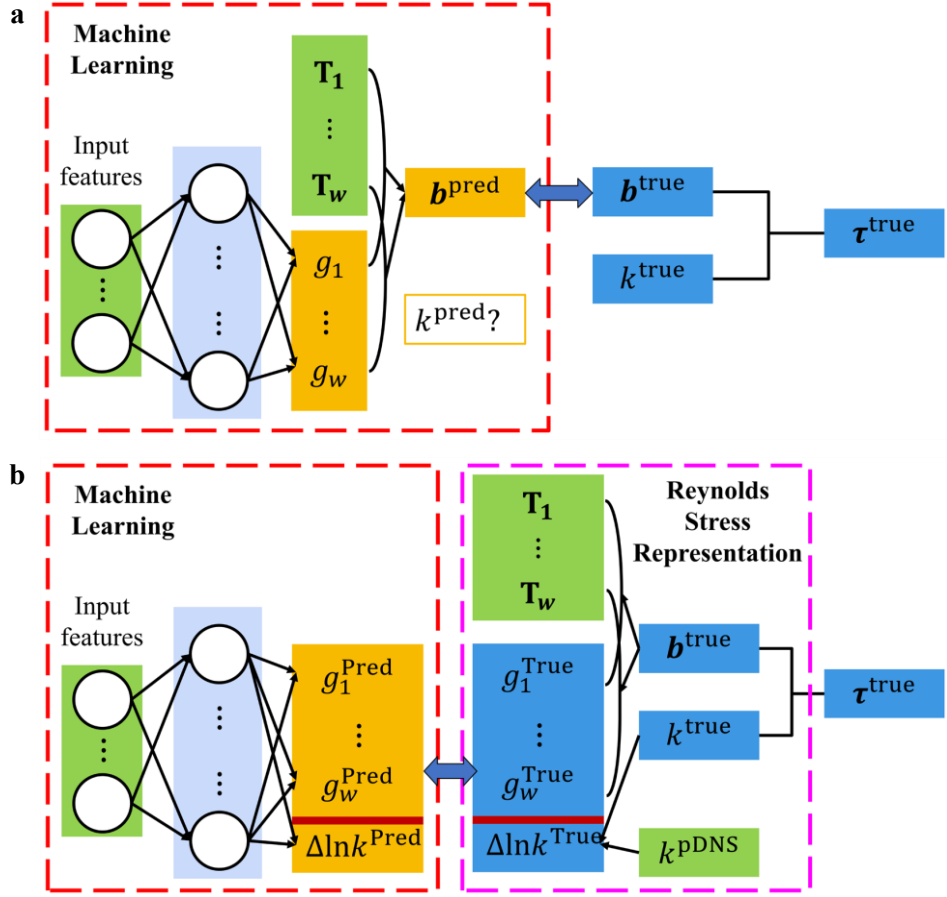

**Fig. 5** Comparison of the entire framework and predicting target selection. **a** Previous framework[13]. **b** Current framework.

### 4. Model training and prediction

The model training and prediction flow chart is shown in **Fig. 6**, which can be summarized in the following steps:

1) Interpolate the DNS mean flow results onto the RANS grid and discretize to acquire $\mathbf{q}^{DNS}$, interpolate the DNS Reynolds stress onto the RANS grid to acquire $\boldsymbol{\tau}^{DNS}$.

2) Iterate the turbulence equations with the mean flow quantities frozen as the DNS results to acquire $(k,\omega)^{pDNS}$.

3) Compute the TKE discrepancy $\Delta \ln k$ and representation coefficients $g_i$ using $\boldsymbol{\tau}^{DNS}$, $\mathbf{q}^{DNS}$, and $(k,\omega)^{pDNS}$.

4) Train the ML model: $f:\{\mathbf{q}^{DNS},(k,\varepsilon)^{pDNS}\}\rightarrow\{\Delta \ln k, g_i\}$.

5) During the application, the computation restarts from the baseline RANS results. In each iteration step, the ML model is executed using $\mathbf{q}$ and $(k,\omega)$ to predict the Reynolds stress $\boldsymbol{\tau}^{pre}$.

6) The final mean flow $\mathbf{q}|_{final}$ is acquired after the computation converges.

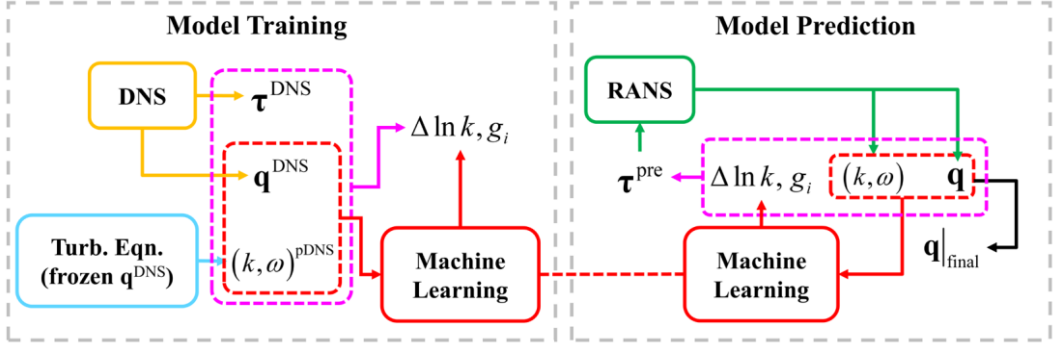

**Fig. 6** Model training and prediction flow chart

## C. Representation coefficient computation

**1. Tensor basis selection**

The 7 tensor bases given in Eqn. (21) are theoretically independent. However, not all bases are selected considering the practical representation effect, numerical error, and computational robustness. The tensor bases are gradually added into the set one by one from low-degree to high-degree. The corresponding representation effect and the mean flow result achieved by substituting the acquired $\{\Delta \ln k, g_i\}$ into the CFD solver are evaluated.

It is found that the addition of the pressure gradient $\mathbf{v}_p$ correlation terms does not significantly improve the representation effect and leads to divergence during the substitution computation. In addition, the high-degree tensor polynomials such as $\mathbf{T}_5$ - $\mathbf{T}_7$ are small-valued and exhibit strong numerical oscillation, which also has no effect on the representation process. Therefore, 3 tensor bases are selected and remarked as $\mathbf{T}_1$ - $\mathbf{T}_3$:

$$\mathbf{T}_1 = \hat{\mathbf{S}}, \quad \mathbf{T}_2 = \hat{\mathbf{S}}\hat{\mathbf{\Omega}} - \hat{\mathbf{\Omega}}\hat{\mathbf{S}}, \quad \mathbf{T}_3 = \hat{\mathbf{v}}_k \otimes \hat{\mathbf{v}}_k \tag{29}$$

The components in the shear stress direction are shown in **Fig. 7**.

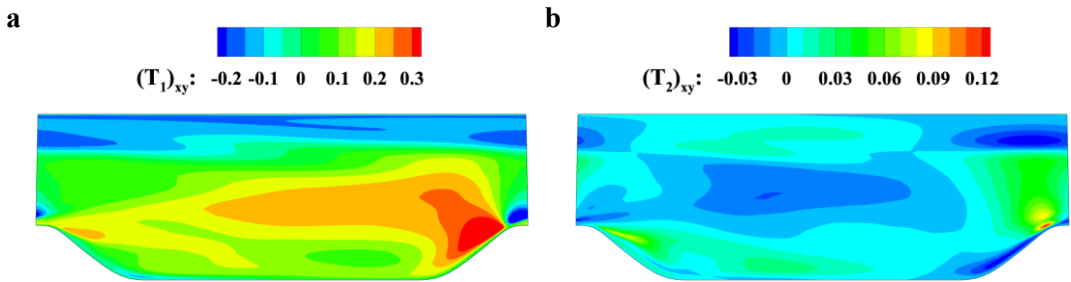

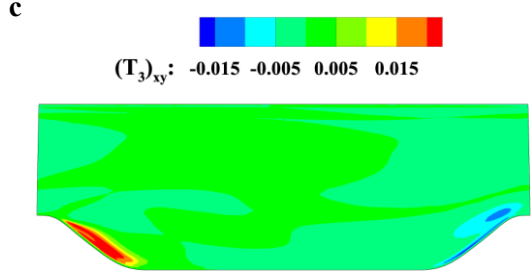

**Fig. 7** The components in the shear stress direction of the selected tensor bases. **a** $T_1 = S$. **b** $T_2 = S\Omega - \Omega S$. **c** $T_3 = v_k \otimes v_k$

**2. Adaptive regularization method**

The representation coefficients can be computed by solving the optimization problem given by Eqn. (27). However, it is found that such acquired coefficients lack smoothness and physical interpretability if the original expression is directly used. Taking the $\alpha = 1.0$ case as an example, the coefficient $g_1$ acquired by directly solving Eqn. (27) is shown in **Fig. 8** (a), and the corresponding Reynolds shear stress is shown in Fig. 9 (a). The true value is shown in Fig. 9 (e). Although the representation effect is quite promising, the coefficient distribution is far from satisfactory. The presented value range is truncated for the following comparison, while the actual value range is far greater. Such severely varying coefficients will result in substantial prediction error during the ML model training.

In addition to the magnitude, the distribution of directly solved coefficients also faces nonphysical and unsmooth problems. The relation between $g_1$ and eddy viscosity $v_t$ can be deduced by taking the first term in Eqn. (15):

$$\tau - \frac{2k}{3}I = -2v_t S = 2k(\mathbf{b}) = 2k\left(g_1 \hat{S}\right) \tag{30}$$

The original expression of the eddy viscosity is $v_t = C_\mu k/\omega$, where $C_\mu$ is an empirical parameter requiring calibration. As it is a constant, $C_\mu$ can be included in the definition of $v_t$, which is the procedure of the program in the current research. Therefore, the equation above is further simplified by substituting the expression of $v_t$:

$$g_1 = -v_t/(k/\omega) = -1 \tag{31}$$

Note that whether $C_\mu$ is included in the definition of $v_t$ varies in different programs. In the current program, $C_\mu$ is included.

Therefore, a negative $g_1$ corresponds to a positive eddy viscosity and positive dissipation.

However, the value of the directly solved $g_1$ is positive in a considerable part of the flow field. The CFD solving process diverges when substituting the coefficients into the RANS equations, which confirms the deterioration of robustness because of negative dissipation. The unsmoothness is marked by the dashed box in **Fig. 8** (a), where the coefficient jump occurs.

In consideration of the coefficient distribution, it is found that the non-physical and unsmooth areas are mainly located at the main flow near the upper surface. The flow field in these areas has almost no mean flow characteristics such as the velocity gradient; therefore, there is almost no turbulence production. However, because of the spatial transport effect of the turbulence, the Reynolds stress can still be conveyed from other areas. If the coefficients are still computed using the original expression (Eqn. (27)) in these areas, the representation process can be analogous to "dividing by zero" and lead to the value jump. Although the Reynolds stress in these areas cannot be accurately represented, the actual magnitude is relatively small, and the influence on the mean flow is limited, which reminds us that the coefficients in these areas can be specifically processed to ensure smoothness.

To overcome the problems above, two methods are proposed in the current research. First, the value range of $g_1$ is constrained to ($-\infty$, 0) to ensure positive dissipation, which can be solved using the constrained least square method.

Second, we introduce the regularization term $R$. Regularization is a commonly employed method to limit the drastic changes of coefficients. By adding an additional term into the target function, the optimization not only minimizes the original target but also considers the effect of $R$. Taking the neural network training as an example, the modified loss function $\tilde{L}$ with the commonly used L2 regularization term is generally defined as:

$$\tilde{L} = L + \lambda \sum_{i=1}^{n} |w_i|^2 \qquad (32)$$

where $w_i$ represents the model weights and $\lambda$ represents the regularization parameter controlling the regularization intensity. The training process minimizes the original loss $L$ and the magnitude of $w_i$ simultaneously, which can avoid overfitting due to large model coefficients.

In the current research, the purpose is to ensure smoothness and computational robustness. As proven above, a small coefficient $g_1$ does not fulfill the requirement. Therefore, physics-informed regularization is introduced, ensuring that the computed representation coefficients do not deviate much from the baseline turbulence model result, which is represented by $g_i^{base}$. The regularization term for each representation coefficient is defined as:

$$R_i = \begin{cases} g_i^2 & \text{, if } g_i^{base} \text{ does not exist} \\ \left(g_i / g_i^{base} - 1\right)^2 & \text{, if } g_i^{base} \text{ exists} \end{cases} \quad (33)$$

Such a definition is compatible with different baseline models. If linear eddy viscosity models are employed, only $g_1^{base}$ exists. If a nonlinear eddy viscosity model is employed, additional prior knowledge can also be included. The optimization problem with regularization is:

$$g_i = \arg\min\left(\|\mathbf{b} - g_i \mathbf{T}_i\|^2 + R\right) = \arg\min\left(\|\mathbf{b} - g_i \mathbf{T}_i\|^2 + \lambda \sum_{i=1}^{w} R_i\right) \quad (34)$$

To verify the regularization effect, large and small $\lambda$ values are selected, and the $g_1$ distributions are shown in **Fig. 8** (b) and (c). The corresponding Reynolds shear stress distributions are shown in **Fig. 9** (b) and (c). The comparison shows that employing regularization surely constrains the coefficient near the baseline value but also decreases the representation effect. $\lambda = 0.1$ leads to better smoothness, but the discrepancy between the true stress and represented stress is larger. In contrast, $\lambda = 0.001$ cannot eliminate the unsmooth area, but the represented stress is closer to the true value.

In summary, employing a unified regularization parameter in the entire flow field cannot meet the requirements of improving the representation effect in key areas and ensuring smoothness in other areas simultaneously. To overcome this deficiency, combined with the previous analysis of the nonphysical and unsmooth problems, an adaptive regularization method based on the magnitude of the tensor basis is proposed in the current research. $\lambda$ varies for different $g_i$ and is defined as:

$$\lambda_i = \lambda_{\min} + \left(\lambda_{\max} - \lambda_{\min}\right)\beta\left(\|\mathbf{T}_i\|\right) \quad (35)$$

where $\lambda_{\min}$ and $\lambda_{\max}$ are the minimum and maximum values of the representation parameter, respectively, $\beta$ is the multiplying factor function of the tensor basis norm $\|\mathbf{T}_i\|$, and the expression is:

$$\beta = \frac{1}{2} - \frac{1}{2}\tanh\frac{2\left(\|\mathbf{T}_i\| - \theta_i\right)}{\theta_i} \quad (36)$$

where $\theta_i$ is the threshold for different $\mathbf{T}_i$, which is predetermined manually based on the distribution of $\|\mathbf{T}_i\|$. When $\|\mathbf{T}_i\| > \theta_i$, $\beta$ approaches 0, and $\lambda$ approaches $\lambda_{\min}$; conversely, when $\|\mathbf{T}_i\| < \theta_i$, $\beta$ approaches 1 and $\lambda$ approaches $\lambda_{\max}$. In the current research, the $\theta_i$ for each $\mathbf{T}_i$ is [0.1, 0.05, 0.01]. $\lambda_{\min}$ and $\lambda_{\max}$ are 0.001 and 0.1, respectively. It is worthy mentioning that the selection of $\theta_i$ seems to be ad hoc. The entire strain rate and rotation rate magnitude will no doubt vary a lot for different flow conditions. But after normalized by the turbulent quantities, the nondimensional $\|\mathbf{T}_i\|$ actually doesn't vary so much. The final prediction results of the stress field and mean flow field at different geometries can prove the validity.

The computed coefficient distribution and corresponding Reynolds shear stress are shown in **Fig. 8** (d) and **Fig. 9** (d). The coefficient distribution and the represented Reynolds shear stress with adaptive regularization show better smoothness in those mainstream areas compared with the small regularization and better accuracy in those critical areas such as the separation shear layer at the same time. Therefore, the results with adaptive regularization are selected as the goal of the following ML training.

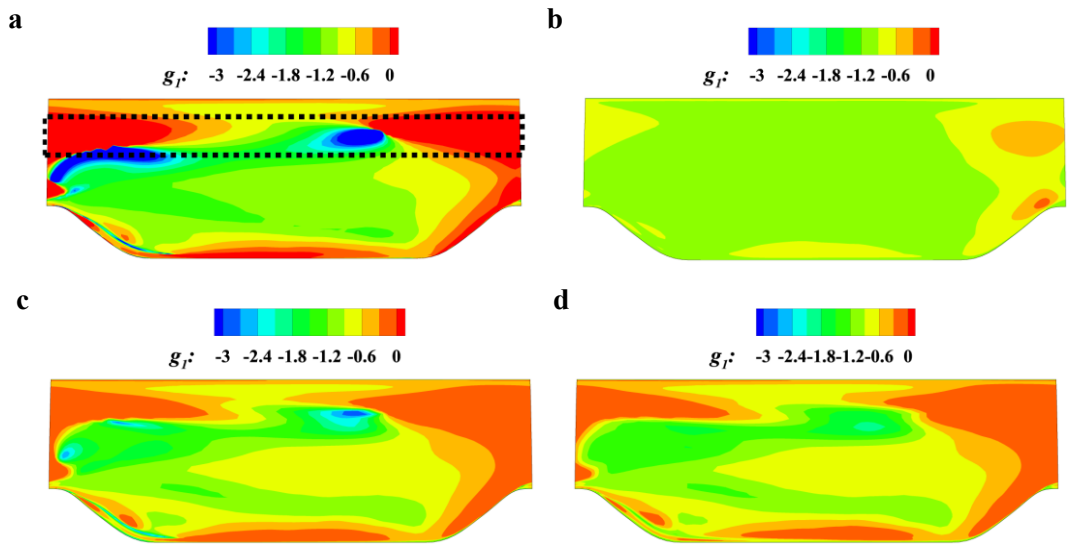

**Fig. 8** Coefficient distribution comparison of different regularizations.  **a** No regularization ($\lambda = 0$). **b** Large regularization ($\lambda = 0.1$). **c** Small regularization ($\lambda = 0.001$). **d** Adaptive regularization (varying $\lambda$)

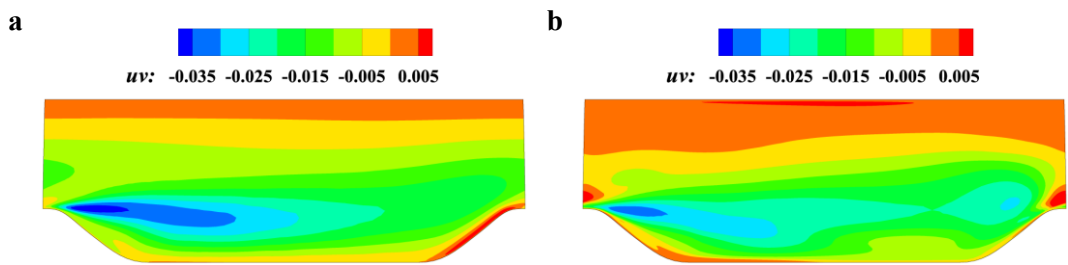

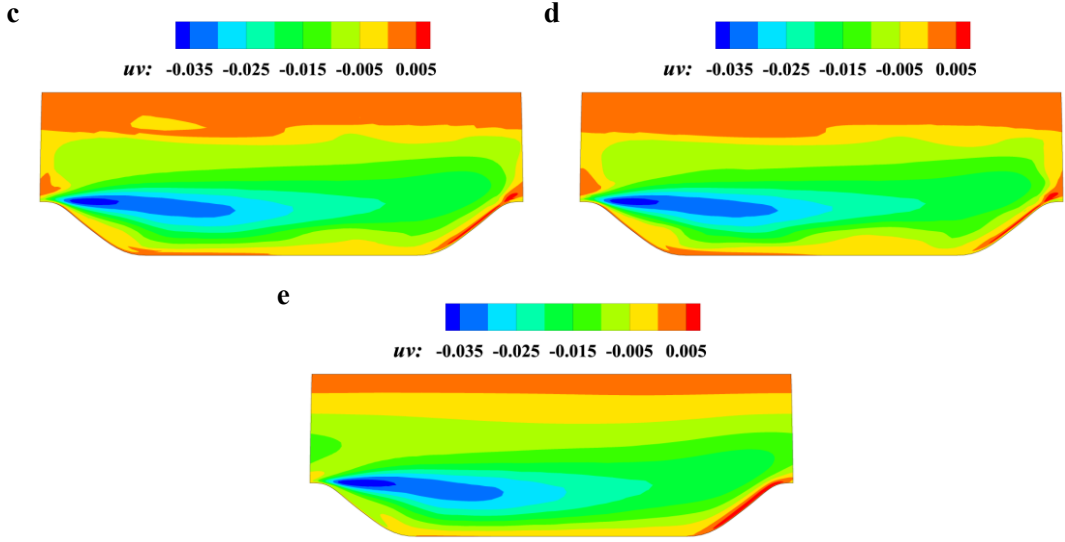

**Fig. 9** Reynolds shear stress distribution comparison of different regularizations. **a** No regularization ($\lambda = 0$). **b** Large regularization ($\lambda = 0.1$). **c** Small regularization ($\lambda = 0.001$). **d** Adaptive regularization (varying $\lambda$). **e** True value.

The final acquired prediction targets $\{\Delta \ln k, g_i\}$ are shown in **Fig. 10** (a), (c), (e), and (g). The corresponding Reynolds shear stress represented by each tensor basis is shown in **Fig. 10** (b), (d), and (f). The first tensor basis $\mathbf{T}_1 = \mathbf{S}$, corresponding to the linear eddy viscosity part, still occupies the majority of the entire representation result. $\mathbf{T}_2 = (\mathbf{S\Omega} - \mathbf{\Omega S})$ and $\mathbf{T}_3 = (\mathbf{v}_k \otimes \mathbf{v}_k)$ supplement the Reynolds stress near the separation shear layer and the boundary layer near the curved wall. The $\Delta \ln k$ distribution indicates that the TKE is underestimated near the separation shear layer, which is an important reason for the inaccurate separation bubble prediction.

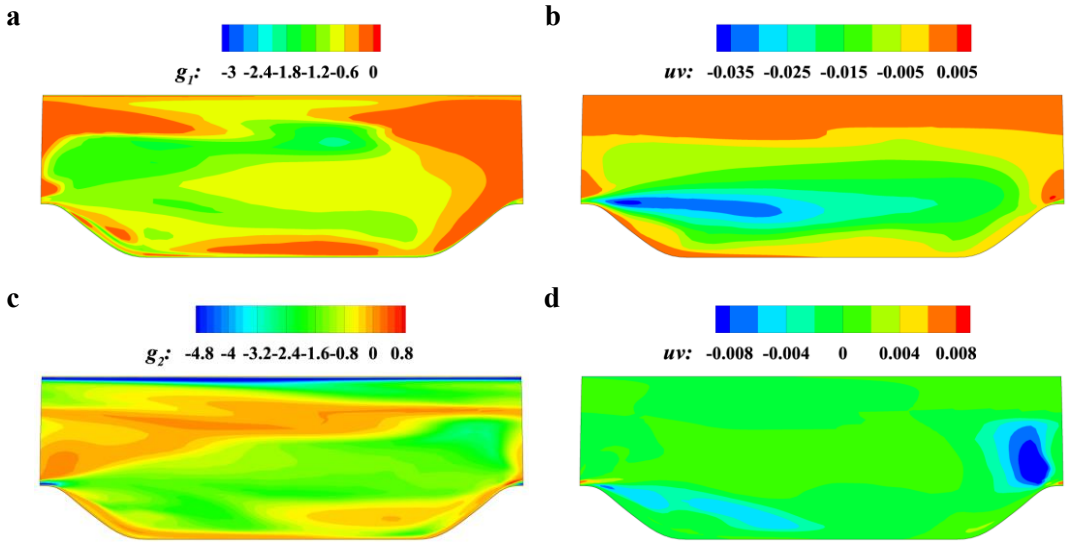

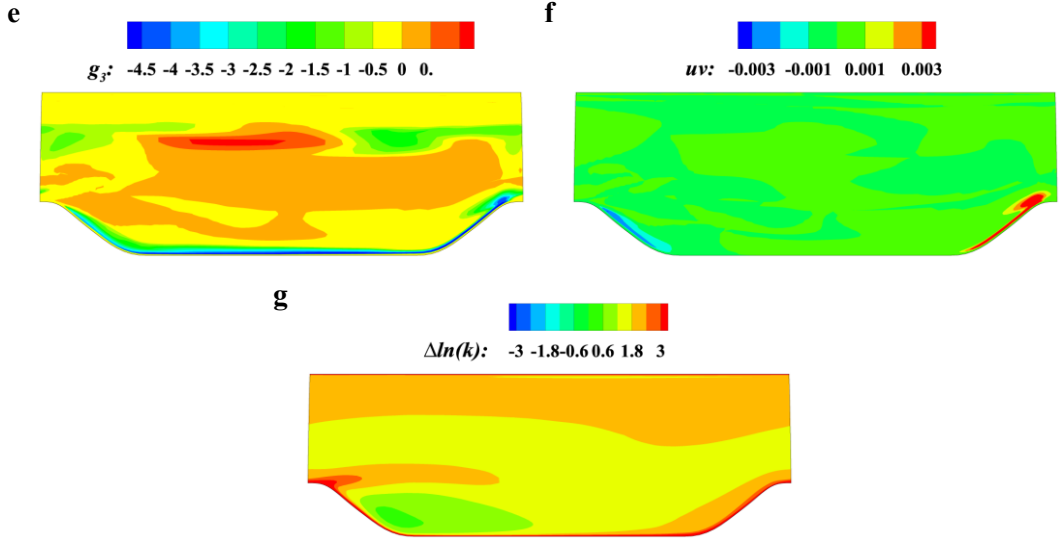

**Fig. 10** Predicting targets and corresponding Reynolds shear stress components. **a** Coefficient $g_1$. **b** Shear stress of $g_1 \mathbf{S}$. **c** Coefficient $g_2$. **d** Shear stress of $g_2(\mathbf{S\Omega} - \mathbf{\Omega S})$. **e** Coefficient $g_3$. **f** Shear stress of $g_3(\mathbf{v}_k \otimes \mathbf{v}_k)$. **g** $\Delta \ln k$.

## III. Results

### A. ML model training

The ML model is trained after the training data preparation and representation coefficient computation according to the methods above. The random forest (RF) is selected as the ML model in the current research. The RF model is a representative of integrated learning that is realized by combining multiple simple learning machines. During the training of RF, the entire training set is separated into several subsets and a decision tree is trained on each subset. Random feature selection is introduced to ensure the "good but different" characteristics of the decision trees and avoid overfitting. The final prediction result is acquired by averaging all the decision tree results. The entire flow chart is shown in **Fig. 11**. The ML training is realized using the sci-kit learn package[48] in Python.

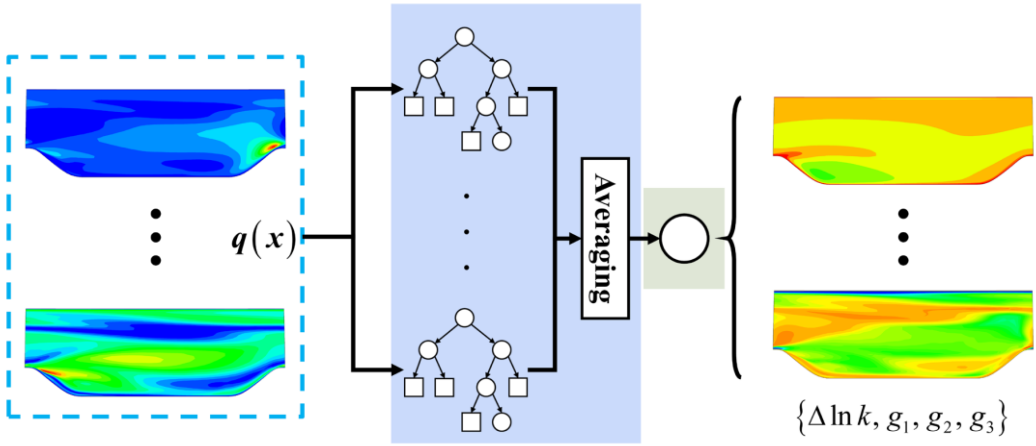

**Fig. 11** The entire framework of RF

Four RF models are trained for each target. The number of decision trees is 500. The loss function type is the mean absolute error. The maximum feature number is $\log_2(9) \approx 3$ according to the "log2" rule. The feature importance rankings are shown in **Fig. 12**. The influence of features differs in each target, indicating the rationality of the individual training of each model.

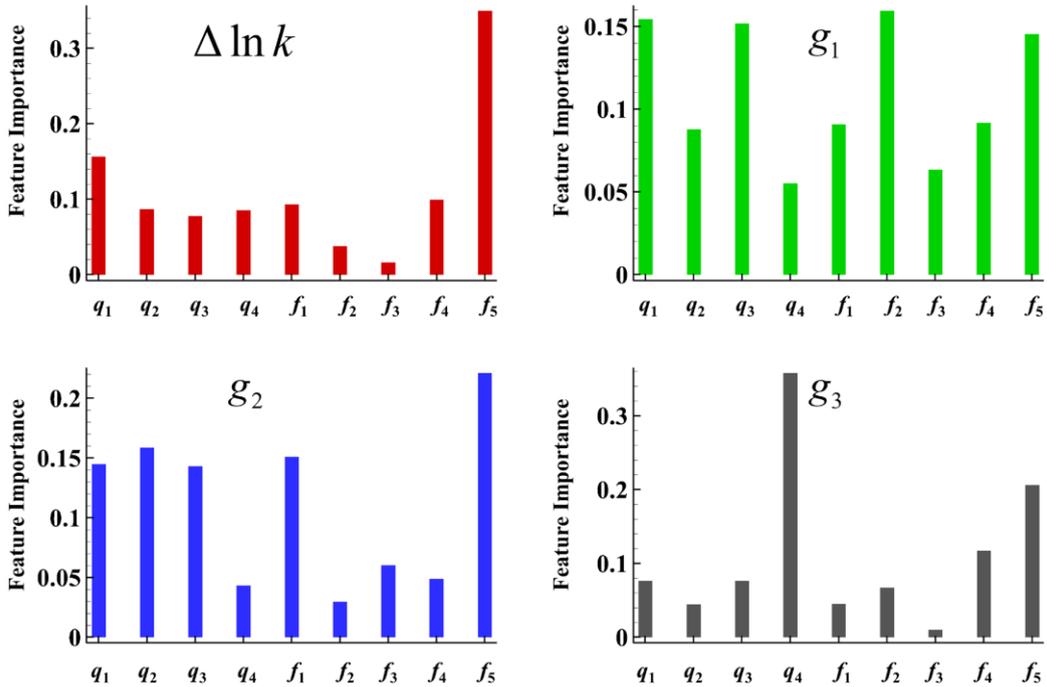

**Fig. 12** Feature importance rankings of different targets

### B. Iterative computation platform

The trained ML model needs to be coupled with the CFD code and executed in each step. The CFD code used in the current research is the open-source CFL3D version 6.7 programmed by Fortran, while the ML model is constructed in Python. The solving process will be extremely inefficient if the data are transferred by writing/reading files. Therefore, an iterative computation platform based on hybrid programming of Fortran/Python is developed. With the help of the CFFI module in Python, the Python part is packaged into a dynamic link library and coupled with CFL3D. The flow chart of computation is shown in **Fig. 13**.

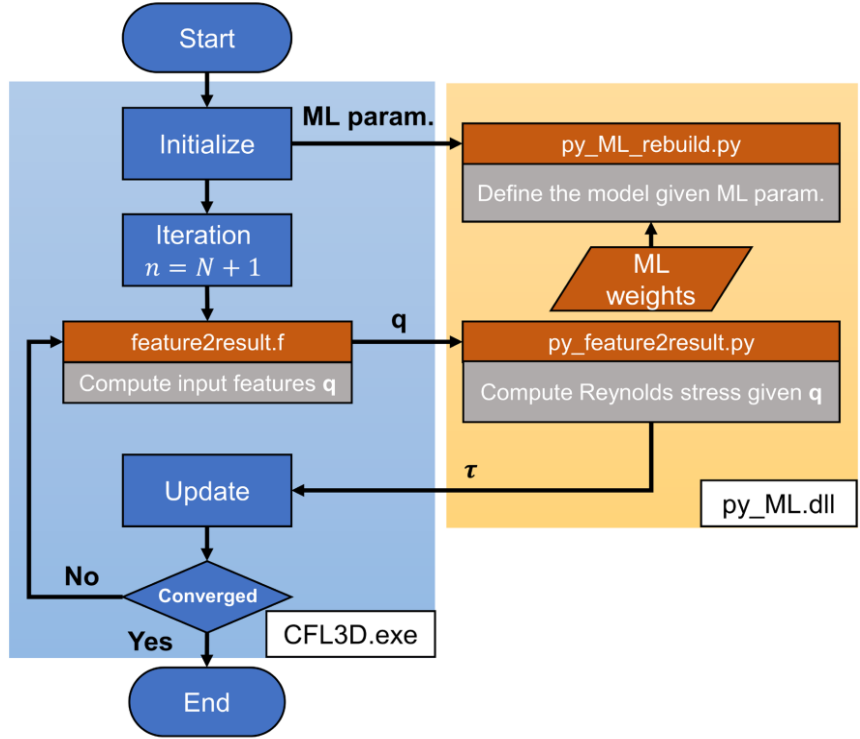

**Fig. 13** Iterative computation platform flow chart

### C. Prediction performance

The Reynolds stress and mean flow prediction are evaluated after the iterative computation converges. To verify the generalization capability, we mainly focus on the performance of the three testing sets. From the aspect of the hill slope, the $\alpha = 0.5$ and 1.5 cases are extrapolation cases, and the $\alpha = 1.0$ case is an interpolation case. The comparison between the predicted and theoretical values of Reynolds shear stress for the testing cases is shown in **Fig. 14**. The theoretical values are computed by the Reynolds stress representation using the true Reynolds stress and mean flow field. Therefore, the theoretical value is only the part that can be represented by three tensor bases but is close enough to the true stress, as compared in **Fig. 9**. More details about the treatment of the discrepancy between the two distributions are discussed in Section IV.

The theoretical values indicate that the Reynolds stress of the testing cases are quite different in distribution and magnitude, which requires the ML model to have a strong generalization capability. The prediction values are close to the theoretical values in key areas, such as the separated shear layer and boundary layer, but there still exist some unsmooth areas.

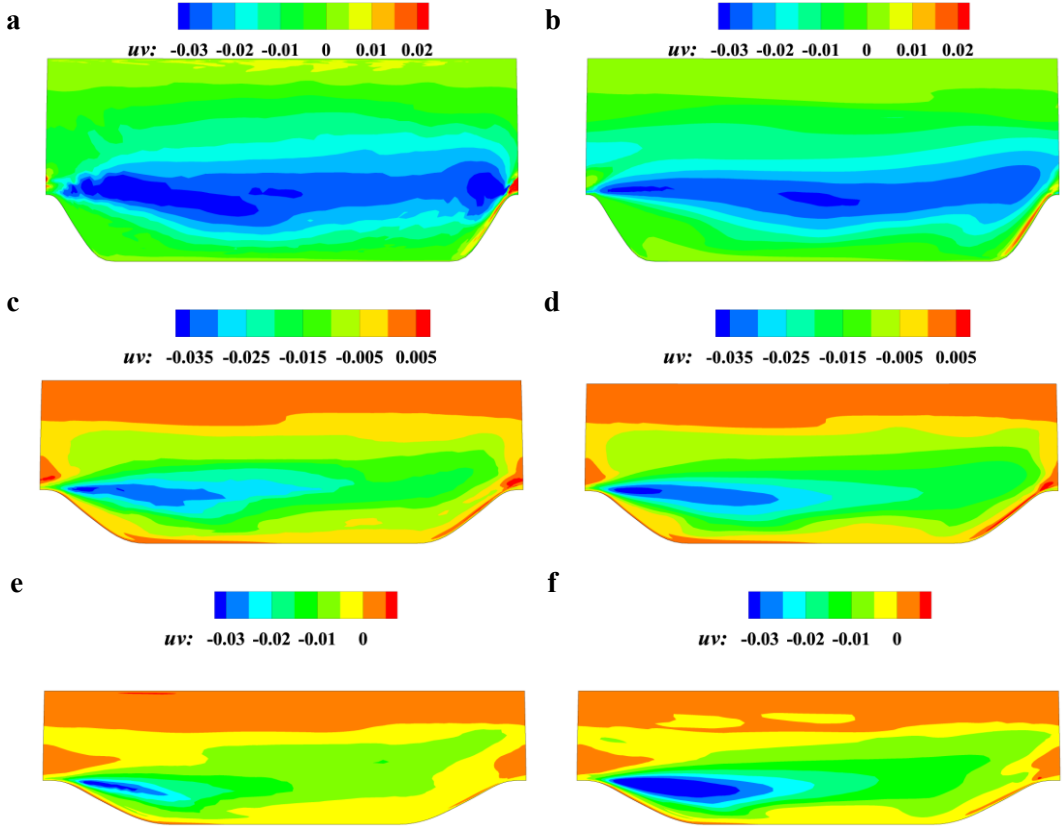

**Fig. 14** Reynolds stress comparison of testing sets. **a** $\alpha = 0.5$ (extrapolation) prediction. **b** corresponding theoretical value. **c** $\alpha = 1.0$ (interpolation) prediction. **d** corresponding theoretical value. **e** $\alpha = 1.5$ (extrapolation) prediction. **f** corresponding theoretical value.

Further evaluation of the mean flow field. The velocity contours are shown in **Fig. 15**, and the profiles are shown in **Fig. 16**. The comparison of lower surface friction is shown in **Fig. 17**. In these comparisons, the reference values are selected as the true value directly from the DNS results since the final purpose of the data-driven turbulence modeling is to correct the mean flow field.

The velocity contour comparison indicates that the mean flow field has better smoothness than the Reynolds stress distribution. This is reasonable because the Reynolds stress can be regarded as an external source term in the velocity transport equation. The unsmooth distribution of the Reynolds stress can be modified by the transport and dissipation of the RANS equation. The iterative embedding framework of the ML model can further increase the coupling effect of the mean flow and the Reynolds stress. The unsmoothness problem near the mainstream area in our previous work[32] is also resolved, which confirms the effect of adaptive regularization. In addition to the smoothness, the prediction accuracy of the flow separation and reattachment is also satisfactory, especially in the small slope case ($\alpha = 1.5$), proving the generalization capacity.

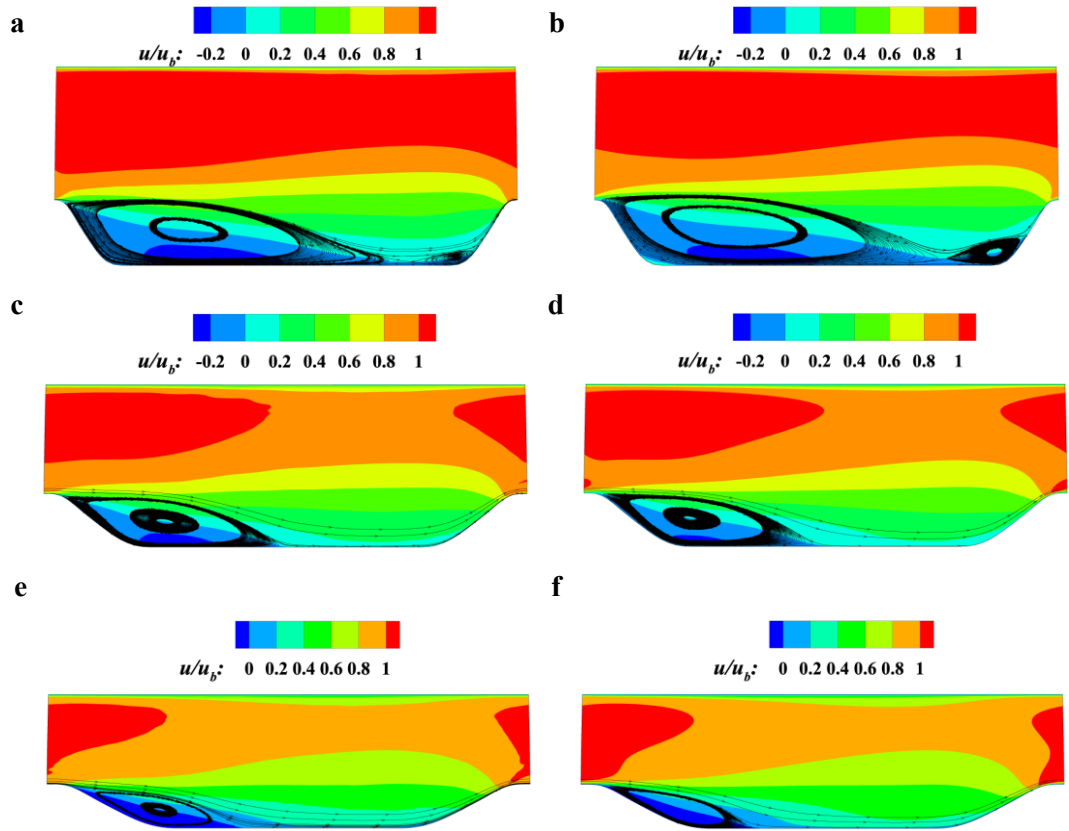

**Fig. 15** Velocity contour comparison of testing sets. **a** $\alpha = 0.5$ (extrapolation) prediction. **b** $\alpha = 0.5$ DNS true value. **c** $\alpha = 1.0$ (interpolation) prediction. **d** $\alpha = 1.0$ DNS true value. **e** $\alpha = 1.5$ (extrapolation) prediction. **f** $\alpha = 1.5$ DNS true value.

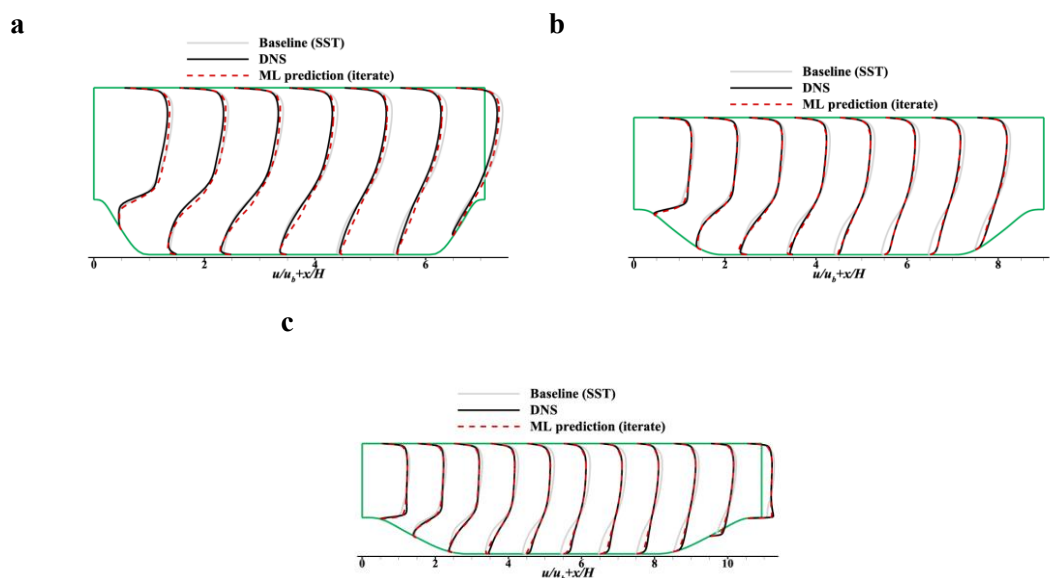

**Fig. 16** Velocity profile comparison of both the training and testing sets. **a** $\alpha = 0.5$ (extrapolation). **b** $\alpha = 1.0$ (interpolation). **c** $\alpha = 1.5$ (extrapolation)

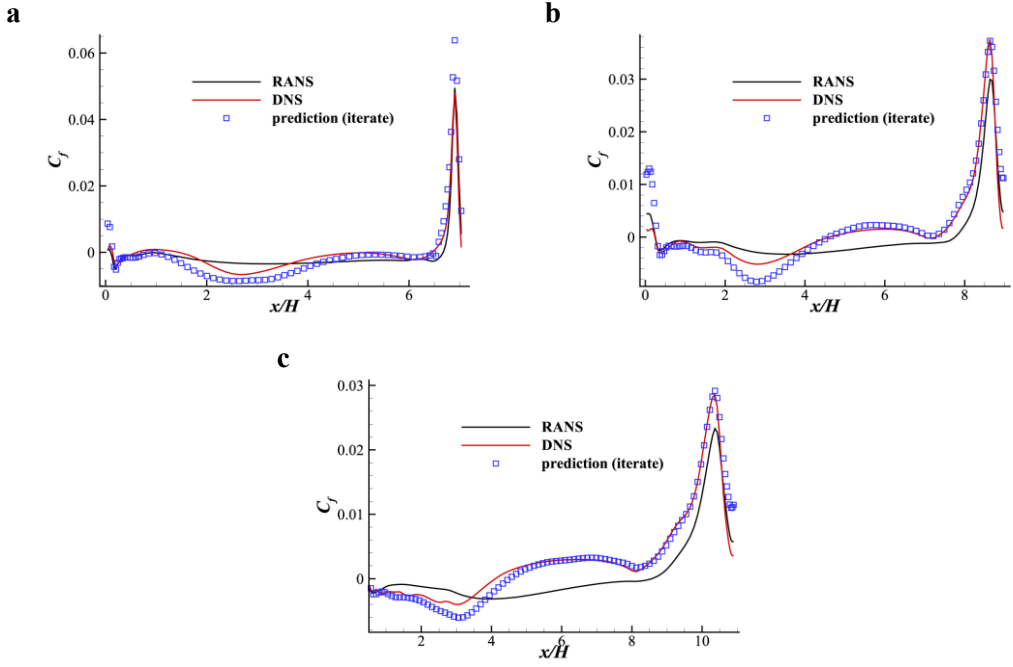

**Fig. 17** Lower surface friction coefficient comparison of both the training and testing sets. **a** $\alpha = 0.5$ (extrapolation). **b** $\alpha = 1.0$ (interpolation). **c** $\alpha = 1.5$ (extrapolation)

## IV. Discussion

The main framework, key methods, and prediction performance are illustrated in the two sections above. Two more issues are discussed in this section. The first concerns the remaining Reynolds stress. Although the discrepancy between the representation stress and the true value stress is small, it still exists. Can the prediction result be further improved if the remaining part is also included in the targets? The second concerns the ML model calling frequency. Because the data are transferred between two different programs, the efficiency still needs further improvement. What if the ML model is not executed in each iteration step, but instead executed after an interval of several steps? The effects are evaluated below.

### A. Effect of the remaining part after the Reynolds stress representation

The remaining part of the Reynolds stress, $\tau_b$, is defined as the discrepancy between the represented Reynolds stress, $\tau_L$, and the true value of Reynolds stress $\tau$:

$$\boldsymbol{\tau}_b = \boldsymbol{\tau} - \boldsymbol{\tau}_L = \boldsymbol{\tau} - 2k\left(\frac{1}{3}\mathbf{I} + g_i \mathbf{T}_i\right) \tag{37}$$

As $\tau$ and $\tau_L$ are second-order symmetric tensors, $\tau_b$ is also a symmetric tensor. Therefore, it can be eigen-decomposed:

$$\tau_b = R\Lambda R^{-1} \tag{38}$$

where $\Lambda = \text{diag}(\lambda_1, \lambda_2, \lambda_3)$ and $R = [v_1, v_2, v_3]$ is the rotation matrix constructed by three eigenvectors. Note that $\lambda_1 + \lambda_2 + \lambda_3 = 0$ because the TKE is included in $\tau_L$., therefore, only two eigenvalues are independent.

To better utilize the represented result, the targets of eigenvalues are defined as the discrepancy between $\tau_L$ and $\tau$:

$$\Delta\lambda_i = \lambda_i - \lambda_i^L, i = 1, 2 \tag{39}$$

The original eigenvectors cannot be directly used as predicting targets because they are not spatially invariant. Therefore, the same process as used in the previous research[3,32] is needed, which is to compute the Euler angles describing the rotation from $\tau_L$ to $\tau$. The details can be found in the mentioned literature and are not listed here. A schematic of the process is shown as:

$$\tau_L \xrightarrow{(R_L)^{-1}} \Lambda^L \xrightarrow{\Delta\lambda_i} \Lambda \xrightarrow{R} \tau$$
$$\therefore R_b = R(R_L)^{-1} \tag{40}$$

$$R_b = \begin{bmatrix} \cos(\varphi_3^z) & \sin(\varphi_3^z) & 0 \\ -\sin(\varphi_3^z) & \cos(\varphi_3^z) & 0 \\ 0 & 0 & 1 \end{bmatrix} \begin{bmatrix} \cos(\theta_2^y) & 0 & \sin(\theta_2^y) \\ 0 & 1 & 0 \\ -\sin(\theta_2^y) & 0 & \cos(\theta_2^y) \end{bmatrix} \begin{bmatrix} 1 & 0 & 0 \\ 0 & \cos(\psi_1^x) & \sin(\psi_1^x) \\ 0 & -\sin(\psi_1^x) & \cos(\psi_1^x) \end{bmatrix} \tag{41}$$

where only $\varphi_3^z$ is effective in the two-dimensional case.

In summary, if $\tau_b$ is considered, there are 3 more predicting targets: $\{\Delta\lambda_1, \Delta\lambda_2, \varphi_3^z\}$. To verify the effects of adding these targets, three scenarios are evaluated, and the corresponding predicting targets are listed below:

(1) Not considering $\tau_b$: $\{\Delta\ln k, g_1, g_2, g_3\}$ (reference)

(2) Considering only the eigenvalues: $\{\Delta\ln k, g_1, g_2, g_3, \Delta\lambda_1, \Delta\lambda_2\}$

(3) Considering the eigenvalues and eigenvectors: $\{\Delta\ln k, g_1, g_2, g_3, \Delta\lambda_1, \Delta\lambda_2, \varphi_3^z\}$

The computation of the three scenarios is performed for the $\alpha = 1.0$ case. To eliminate the effect of the ML model prediction error, the targets above are all frozen as the theoretical values during the computation. The mean flow velocity contours are shown in **Fig. 18**. The comparison indicates that the introduction of $\tau_b$ does not improve the prediction and even worsens smoothness and convergence. Therefore, the method of discarding $\tau_b$ employed in the current research is rational.

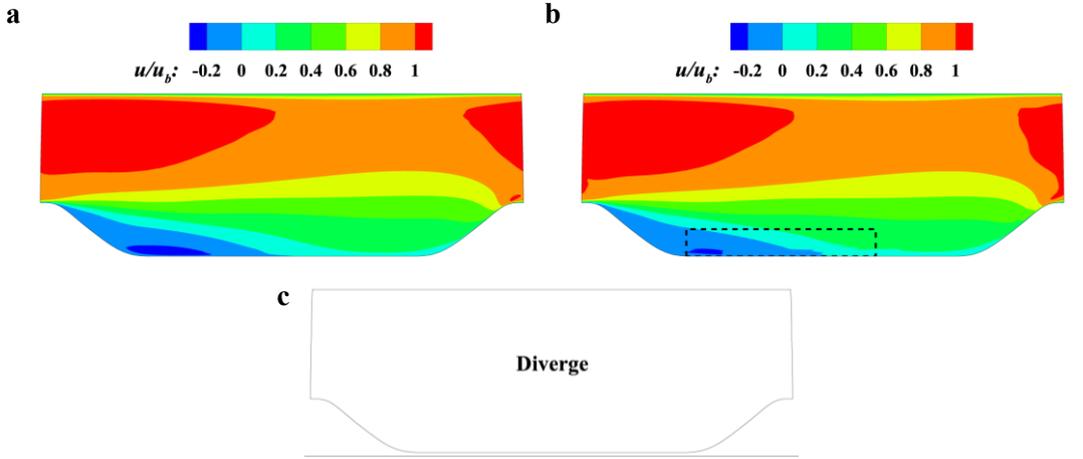

**Fig. 18** Velocity contour comparison of different treatments. **a** Not considering $\tau_b$. **b** Considering only the eigenvalues. **c** Considering the eigenvalues and eigenvectors.

## B. Effect of the ML model calling frequency

As described in **Fig. 3** and **Fig. 13**, the original iterative flow chart needs to call the ML model and update the Reynolds stress in each iteration step. This will increase the time cost significantly compared with the baseline RANS computation. Still taking the $\alpha = 1.0$ case as an example, the grid number of the case is 77 in the normal direction and 89 in the streamwise direction, for a total 6853 of points. The time costs of the direct SST computation and the iterative coupling calculation are listed in **Table 2**. It can be observed that the iteration cost increases considerably, but the time cost relative to unsteady simulations such as LES or DNS is still acceptable.

A natural idea is to execute the ML model after an interval of several steps. To verify the effect, we modify the program and test four intervals: executing the ML model every 1, 3, 5, and 10 steps. The time costs are also listed in **Table 2**. The results show that the computation will diverge if the interval is too large. The ratio of total time cost between different intervals is essentially the same as the ratio of the interval steps, which indicates that the data transfer and Python computation occupy the main proportion of the time cost. The 3 converged mean flow results are generally the same, but the per 5 step case shows vibration near the periodic hill top, as shown in **Fig. 19**.

In summary, executing the ML model after an interval can accelerate the computation. However, the smoothness and the convergence will be affected if the interval step is too large.

**Table 2:** Time cost comparison

| Computation type | Average time cost of each iteration step | Minimum steps needed for convergence | Total time cost |
| --- | --- | --- | --- |
| Direct SST | 0.03s | ~5000 | 150s |
| Executed every 1 step | 2.25s | ~2000 | 4500s |
| Executed every 3 steps | 0.80s | ~2000 | 1600s |
| Executed every 5 steps | 0.48s | ~2000 | 950s |
| Executed every 10 steps | 0.25s | Diverges | / |

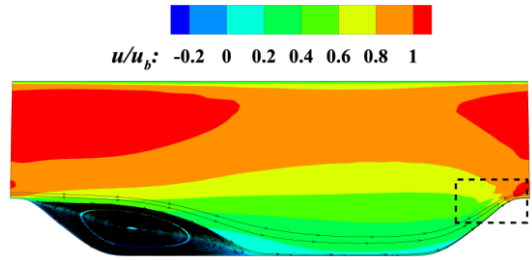

**Fig. 19** Mean flow velocity of the "Executed every 5 steps" case

## V. Conclusion

An iterative data-driven turbulence modeling framework based on Reynolds stress representation is established in the current research. Reynolds stress representation analysis is proposed, including extending the relevant tensor arguments and deducing the complete irreducible tensor invariants and integrity basis. Then, the four steps of the model construction are illustrated. The truth compatibility principle of the training data is proposed. A novel framework separating the Reynolds stress representation and ML is proposed. After the framework construction, the calculation process of the representation coefficient is discussed in detail. The direct solution of the representation coefficient is proven to produce nonphysical and unsmooth distributions, and the adaptive regularization method proposed in the current research can establish a better tradeoff between smoothness and accuracy. The model performance is evaluated in periodic hill cases, and the results indicate that the iterative framework proposed in the current research has better dynamic convergence characteristics and better solution smoothness and accuracy. Finally, the effects of the remaining part after the Reynolds stress representation and the ML model calling frequency are evaluated.

# Appendix: The invariants and symmetric integrity basis of arbitrary numbers of A$_i$, W$_p$, and v$_m$

We suppose that a symmetric tensor **H** is an isotropic tensor function of a set of tensor arguments $(\mathbf{A}_i, \mathbf{W}_p, \mathbf{v}_m), i = 1, \cdots, L, p = 1, \cdots, M, m = 1, \cdots, N$. The complete and irreducible invariants and integrity basis are listed here.

### A. Three-dimensional situation

The invariants are as follows:

$$
\begin{aligned}
\text{invariants:} \quad & \mathbf{v}_m \cdot \mathbf{v}_m, \mathbf{v}_m \cdot \mathbf{v}_n \\
& \text{tr}\mathbf{A}_i, \text{tr}\mathbf{A}_i^2, \text{tr}\mathbf{A}_i^3, \text{tr}\mathbf{A}_i\mathbf{A}_j, \text{tr}\mathbf{A}_i^2\mathbf{A}_j, \text{tr}\mathbf{A}_i\mathbf{A}_j^2, \text{tr}\mathbf{A}_i^2\mathbf{A}_j^2, \text{tr}\mathbf{A}_i\mathbf{A}_j\mathbf{A}_k, \\
& \text{tr}\mathbf{W}_p^2, \text{tr}\mathbf{W}_p\mathbf{W}_q, \text{tr}\mathbf{W}_p\mathbf{W}_q\mathbf{W}_r, \\
& \mathbf{v}_m \cdot \mathbf{A}_i \mathbf{v}_m, \mathbf{v}_m \cdot \mathbf{A}_i^2 \mathbf{v}_m, \mathbf{v}_m \cdot \mathbf{A}_i\mathbf{A}_j \mathbf{v}_m, \\
& \mathbf{v}_m \cdot \mathbf{A}_i \mathbf{v}_n, \mathbf{v}_m \cdot \mathbf{A}_i^2 \mathbf{v}_n, \mathbf{v}_m \cdot \left(\mathbf{A}_i\mathbf{A}_j - \mathbf{A}_j\mathbf{A}_i\right) \mathbf{v}_n, \\
& \mathbf{v}_m \cdot \mathbf{W}_p^2 \mathbf{v}_m, \mathbf{v}_m \cdot \mathbf{W}_p\mathbf{W}_q \mathbf{v}_m, \mathbf{v}_m \cdot \mathbf{W}_p^2\mathbf{W}_q \mathbf{v}_m, \mathbf{v}_m \cdot \mathbf{W}_p\mathbf{W}_q^2 \mathbf{v}_m, \\
& \mathbf{v}_m \cdot \mathbf{W}_p \mathbf{v}_n, \mathbf{v}_m \cdot \mathbf{W}_p^2 \mathbf{v}_n, \mathbf{v}_m \cdot \left(\mathbf{W}_p\mathbf{W}_q - \mathbf{W}_q\mathbf{W}_p\right) \mathbf{v}_n, \\
& \text{tr}\mathbf{A}_i\mathbf{W}_p^2, \text{tr}\mathbf{A}_i^2\mathbf{W}_p^2, \text{tr}\mathbf{A}_i^2\mathbf{W}_p^2\mathbf{A}_i\mathbf{W}_p, \text{tr}\mathbf{A}_i\mathbf{W}_p\mathbf{W}_q, \text{tr}\mathbf{A}_i\mathbf{W}_p\mathbf{W}_q^2, \text{tr}\mathbf{A}_i\mathbf{W}_p^2\mathbf{W}_q, \\
& \text{tr}\mathbf{A}_i\mathbf{A}_j\mathbf{W}_p, \text{tr}\mathbf{A}_i\mathbf{A}_j\mathbf{W}_p^2, \text{tr}\mathbf{A}_i\mathbf{W}_p^2\mathbf{A}_j\mathbf{W}_p, \text{tr}\mathbf{A}_i\mathbf{A}_j^2\mathbf{W}_p, \text{tr}\mathbf{A}_i^2\mathbf{A}_j\mathbf{W}_p \\
& \mathbf{v}_m \cdot \mathbf{A}_i\mathbf{W}_p \mathbf{v}_m, \mathbf{v}_m \cdot \mathbf{A}_i\mathbf{W}_p^2 \mathbf{v}_m, \mathbf{v}_m \cdot \mathbf{A}_i^2\mathbf{W}_p \mathbf{v}_m, \mathbf{v}_m \cdot \mathbf{W}_p\mathbf{A}_i\mathbf{W}_p^2 \mathbf{v}_m, \\
& \mathbf{v}_m \cdot \left(\mathbf{A}_i\mathbf{W}_p + \mathbf{W}_p\mathbf{A}_i\right) \mathbf{v}_n
\end{aligned} \quad (42)
$$

where $i, j = 1, \cdots, L$ with $i < j$, $p, q = 1, \cdots, M$ with $p < q$, and $m, n = 1, \cdots, N$ with $m < n$.

The integrity basis is as follows:

$$\begin{aligned}
\text{integrity basis} \quad & \mathbf{v}_m \otimes \mathbf{v}_m, \; \mathbf{v}_m \otimes \mathbf{A}_i \mathbf{v}_m + \mathbf{A}_i \mathbf{v}_m \otimes \mathbf{v}_m, \; \mathbf{v}_m \otimes \mathbf{A}_i^2 \mathbf{v}_m + \mathbf{A}_i^2 \mathbf{v}_m \otimes \mathbf{v}_m, \\
\text{of } \mathbf{H}: \quad & \mathbf{I}, \mathbf{A}_i, \mathbf{A}_i^2, \mathbf{A}_i \mathbf{A}_j + \mathbf{A}_j \mathbf{A}_i, \mathbf{A}_i^2 \mathbf{A}_j + \mathbf{A}_j \mathbf{A}_i^2, \mathbf{A}_i \mathbf{A}_j^2 + \mathbf{A}_j^2 \mathbf{A}_i, \\
& \mathbf{W}_p^2, \mathbf{W}_p \mathbf{W}_q + \mathbf{W}_q \mathbf{W}_p, \mathbf{W}_p^2 \mathbf{W}_q - \mathbf{W}_q \mathbf{W}_p^2, \mathbf{W}_p \mathbf{W}_q^2 - \mathbf{W}_q^2 \mathbf{W}_p, \\
& \mathbf{v}_m \otimes \mathbf{W}_p \mathbf{v}_m + \mathbf{W}_p \mathbf{v}_m \otimes \mathbf{v}_m, \mathbf{W}_p \mathbf{v}_m \otimes \mathbf{W}_p \mathbf{v}_m, \\
& \mathbf{W}_p \mathbf{v}_m \otimes \mathbf{W}_p^2 \mathbf{v}_m + \mathbf{W}_p^2 \mathbf{v}_m \otimes \mathbf{W}_p \mathbf{v}_m, \\
& \mathbf{A}_i \mathbf{W}_p - \mathbf{W}_p \mathbf{A}_i, \mathbf{W}_p \mathbf{A}_i \mathbf{W}_p, \mathbf{A}_i^2 \mathbf{W}_p - \mathbf{W}_p \mathbf{A}_i^2, \mathbf{W}_p \left( \mathbf{A}_i \mathbf{W}_p - \mathbf{W}_p \mathbf{A}_i \right) \mathbf{W}_p \\
& \mathbf{v}_m \otimes \mathbf{v}_n + \mathbf{v}_n \otimes \mathbf{v}_m, \mathbf{A}_i \boldsymbol{\Sigma} - \boldsymbol{\Sigma} \mathbf{A}_i, \mathbf{W}_p \boldsymbol{\Sigma} + \boldsymbol{\Sigma} \mathbf{W}_p
\end{aligned} \quad (43)$$

where $\boldsymbol{\Sigma} = \mathbf{v}_m \otimes \mathbf{v}_n - \mathbf{v}_n \otimes \mathbf{v}_m$

## B. Two-dimensional situation

The invariants are as follows:

$$\begin{aligned}
\text{invariants:} \quad & \mathbf{v}_m \cdot \mathbf{v}_m, \; \mathbf{v}_m \cdot \mathbf{v}_n \\
& \text{tr} \mathbf{A}_i, \text{tr} \mathbf{A}_i^2, \text{tr} \mathbf{A}_i \mathbf{A}_j, \\
& \text{tr} \mathbf{W}_p^2, \text{tr} \mathbf{W}_p \mathbf{W}_q, \\
& \mathbf{v}_m \cdot \mathbf{A}_i \mathbf{v}_m, \mathbf{v}_m \cdot \mathbf{A}_i \mathbf{v}_n, \\
& \mathbf{v}_m \cdot \mathbf{W}_p \mathbf{v}_n, \\
& \text{tr} \mathbf{A}_i \mathbf{A}_j \mathbf{W}_p, \\
& \mathbf{v}_m \cdot \mathbf{A}_i \mathbf{W}_p \mathbf{v}_m
\end{aligned} \quad (44)$$

The integrity basis is as follows:

$$\begin{aligned}
\text{integrity basis} \quad & \mathbf{v}_m \otimes \mathbf{v}_m, \\
\text{of } \mathbf{H}: \quad & \mathbf{I}, \mathbf{A}_i, \\
& \mathbf{v}_m \otimes \mathbf{W}_p \mathbf{v}_m + \mathbf{W}_p \mathbf{v}_m \otimes \mathbf{v}_m, \\
& \mathbf{A}_i \mathbf{W}_p - \mathbf{W}_p \mathbf{A}_i, \\
& \mathbf{v}_m \otimes \mathbf{v}_n + \mathbf{v}_n \otimes \mathbf{v}_m
\end{aligned} \quad (45)$$

As a verification, if we take $(\mathbf{S}, \boldsymbol{\Omega})$ as the arguments, the invariants and integrity basis acquired from (44) and (45) are $\{\mathbf{S}^2\}$, $\{\boldsymbol{\Omega}^2\}$ and $\{\mathbf{I}, \mathbf{S}, \mathbf{S}\boldsymbol{\Omega}\text{-}\boldsymbol{\Omega}\mathbf{S}\}$, which is the same as Pope's conclusion in a two-dimensional situation[35].

## Acknowledgments

This work was supported by the National Natural Science Foundation of China (91852108, 11872230, and 92152301).

All the authors contributed equally to this work.

## Data availability statements

The data that support the findings of this study (the dataset of flows over periodic hills of parameterized geometries) are available within the article[45].